\documentclass[letterpaper,twocolumn,american,english,aps,prb]{revtex4}
\usepackage[T1]{fontenc}
\usepackage[latin9]{inputenc}
\usepackage{color}
\usepackage{babel}
\usepackage{float}
\usepackage{bm}
\usepackage{amsmath}
\usepackage{amssymb}
\usepackage{graphicx}
\usepackage{esint}
\usepackage[unicode=true,pdfusetitle,
 bookmarks=true,bookmarksnumbered=false,bookmarksopen=false,
 breaklinks=false,pdfborder={0 0 0},pdfborderstyle={},backref=false,colorlinks=true]
 {hyperref}
\hypersetup{
 pdfborderstyle={},pdfborderstyle={},pdfborderstyle={}}
\usepackage{breakurl}

\makeatletter

\@ifundefined{textcolor}{}
{%
 \definecolor{BLACK}{gray}{0}
 \definecolor{WHITE}{gray}{1}
 \definecolor{RED}{rgb}{1,0,0}
 \definecolor{GREEN}{rgb}{0,1,0}
 \definecolor{BLUE}{rgb}{0,0,1}
 \definecolor{CYAN}{cmyk}{1,0,0,0}
 \definecolor{MAGENTA}{cmyk}{0,1,0,0}
 \definecolor{YELLOW}{cmyk}{0,0,1,0}
}

\usepackage{babel}
\usepackage{bm}

\usepackage{breakurl}

\@ifundefined{textcolor}{}{%
 \definecolor{BLACK}{gray}{0}
 \definecolor{WHITE}{gray}{1}
 \definecolor{RED}{rgb}{1,0,0}
 \definecolor{GREEN}{rgb}{0,1,0}
 \definecolor{BLUE}{rgb}{0,0,1}
 \definecolor{CYAN}{cmyk}{1,0,0,0}
 \definecolor{MAGENTA}{cmyk}{0,1,0,0}
 \definecolor{YELLOW}{cmyk}{0,0,1,0}
}

\usepackage{babel}
\usepackage{dsfont}

\makeatother

\begin{document}

\title{Ferromagnetic resonance of a $2D$ array of nanomagnets: effects
of surface anisotropy and dipolar interactions}

\author{J.-L. Déjardin$^{1}$}
\email{dejardin@univ-perp.fr}

\author{A. Franco$^{2,3}$}
\email{andres.franco@usm.cl}

\author{F. Vernay$^{1}$}
\email{francois.vernay@univ-perp.fr}

\author{H. Kachkachi$^{1}$}
\email{hamid.kachkachi@univ-perp.fr}

\affiliation{$^{1}$Laboratoire PROMES-CNRS (UPR-8521), Université de Perpignan
Via Domitia, Rambla de la Thermodynamique, Tecnosud, 66100 Perpignan,
FRANCE.}

\affiliation{$^{2}$Departamento de Física, Universidad Técnica Federico Santa
María, Avenida España 1680, 2390123 Valparaiso, CHILE\\
 $^{3}$Núcleo de Matemáticas, Física y Estadística, Facultad de Ciencias,
Universidad Mayor, Av. Manuel Montt 367, Providencia, Santiago, CHILE}

\date{\today}
\begin{abstract}
We develop an analytical approach for studying the FMR frequency shift
due to dipolar interactions and surface effects in two-dimensional
arrays of nanomagnets with (effective) uniaxial anisotropy along the
magnetic field. For this we build a general formalism on the basis
of perturbation theory that applies to dilute assemblies but which
goes beyond the point-dipole approximation as it takes account of
the size and shape of the nano-elements, in addition to their separation
and spatial arrangement. The contribution to the frequency shift due
to the shape and size of the nano-elements has been obtained in terms
of their aspect ratio, their separation and the lattice geometry.
We have also varied the size of the array itself and compared the
results with a semi-analytical model and reached an agreement that
improves as the size of the array increases. We find that the red-shift
of the ferromagnetic resonance due to dipolar interactions decreases
for smaller arrays. Surface effects may induce either a blue-shift
or a red-shift of the FMR frequency, depending on the crystal and
magnetic properties of the nano-elements themselves. In particular,
some configurations of the nano-elements assemblies may lead to a
full compensation between surface effects and dipole interactions. 
\end{abstract}
\maketitle

\section{Introduction and statement of the problem}

Today there are various sophisticated ways of fabricating and characterizing
arrays of magnetic nano-elements of tunable magnetic properties that
are of interest in practical applications such as magnetic recording,
hyperthermia, catalysis and so on. In fundamental and theoretical
research, these achievements are welcome as they meet a long-standing
demand for well-defined structures with controllable parameters such
as the size, the shape and spatial organization. On the other hand,
experimental techniques of characterization and measurements have
known great progress with regards to spatial and temporal resolution,
thus further bridging the gap between the nanometer and macroscopic
scales. Ferromagnetic resonance (FMR)\citep{vonsovskii66pp,gurmel96crcpress,tran06phd,heinrich94springer,leeetal11apl,goncalvesetal13apl,Schoeppneretal14jap,ollefsetal15jap}
is one of such very precise techniques that has been upgraded to detect
the resonance of small arrays of nanocubes with a sensitivity of $10^{6}\mu_{\mathrm{B}}$.
Other variants of the FMR spectroscopy, such as the so-called Magnetic
Resonance Force Microscopy (MRFM)\citep{Sidles_RevModPhys} may be
used for the characterization of cobalt nano-spheres \citep{lavenantetal2014nanofab}.
Standard FMR theory \citep{kittel48pr}, based on microwave absorption
in magnetic materials, shows that the resonance frequency is, to a
first approximation, a function of the effective field which usually
comprises the magneto-crystalline and shape anisotropy, the exchange
coupling and the applied (static) magnetic field. This dependence
can be used to characterize the material's parameters. On the other
hand, these parameters can be varied so as to control the microwave
absorption properties of the material. For instance, the dipolar interactions
(DI) between nano-elements can be modulated by the elements density.
A dipolar coupling between (parallel) elongated objects induces an
additional anisotropy with an easy axis along the DI bond\emph{ }and
perpendicular to the objects. In Refs. \onlinecite{strijkersetal99jap,ancinasetal01jap}
it was shown that the direction of the effective anisotropy can be
tuned parallel or perpendicular to the nano-elements axes by varying
the concentration. This two-way relationship between the FMR characteristics
and the system's physical parameters is usually based on analytical
expressions that provide the resonance frequency as a function of
the material's parameters (anisotropy constants, exchange and dipolar
couplings). In the case of an array of interacting magnetic nano-elements
such analytical expressions cannot be obtained in a closed form and
one has to resort to some approximation, \emph{e.g.} that of weak
interactions (which can experimentally be tuned, for instance, for
core-shell nanoparticles \citep{Yang_etal_APL2009}), or equivalently
of dilute assemblies. Accordingly, one can apply perturbation theory
and derive approximate expressions for the resonance frequency of
the interacting assembly, taking into account the size and form of
both the nano-elements and the array, in addition to the (effective)
anisotropy and applied DC field. As the size decreases surface effects
(SE) start to play a critical role in the magnetic properties of the
nanomagnets, especially in monodisperse assemblies with oriented effective
anisotropy. For ratios of the surface anisotropy constant $K_{s}$
to the exchange coupling $J$ smaller than unity ($K_{s}/J<1$) \citep{urquhartetal88jap,skocoe99iop,perrai05springer},
the spin configuration within the nano-magnet may be considered as
quasi-collinear\citep{kacbon06prb, yanesetal07prb}. Then, an effective
model for the (macroscopic) net magnetic moment of the nano-magnet
can be built and which properly accounts for the magnetic properties
(static and dynamical) of the nano-magnet \citep{garkac03prl, kachkachi07j3m, kacbon06prb, yanesetal07prb}.
Using this model, we extend our analytical study by including the
effects of both dipolar interactions and surface anisotropy and their
competition {[}see Section \ref{sec:Surface-effects}{]}. Therefore,
the main objective of the present work is to i) derive the correction
to the resonance frequency due to DI using perturbation theory, beyond
the point-dipole approximation, \emph{i.e.} taking account of the
shape and size of the nano-elements (or dipoles) and ii) derive the
shift of the resonance frequency due to surface effects using the
effective model for each nano-element \citep{kacbon06prb,yanesetal07prb}.
Then, we apply this formalism to the prototypical case of an array
of thin disks and derive the corresponding approximate expression
for the frequency shift induced by DI. Next, we analyze the contribution
from surface anisotropy to the FMR frequency and compare with the
DI-induced shift.

Another general approach has been developed in Ref. \onlinecite{verbaetal12prb}
for studying the collective dipolar (or magnonic) spin-wave excitations
in a two dimensional array of magnetic nano-dots, in the absence of
magneto-crystalline anisotropy. Other similar works and approaches
can be found in the literature \citep{beleggiaetal04jmmm278,naletovetal11prb,sukhovetal14ieee}
which deal with collective effects in assemblies of nano-elements.
In the present work we adopt a general but simple approach that allows
us to take into account surface effects as well as dipolar interactions
and to study their competition, in dilute and monodisperse assemblies
with oriented magneto-crystalline (effective) uniaxial anisotropy.
Moreover, for practical reasons related with the possibility to compare
with experiments, we focus on FMR resonance and provide explicit analytical
expressions for the frequency shift induced by dipolar interactions
and surface anisotropy. As was discussed above, today several experimental
groups\citep{beleggiaetal04jmmm278,shendruketal07nanotech,leeetal11apl,naletovetal11prb,mitsuzukaetal12apl,goncalvesetal13apl,lisnak13icfpm8,sukhovetal14ieee,lavenantetal2014nanofab,khusrhidetal14jap,lavoratoetal15jpcc,ollefsetal15jap}
are able to fabricate well organized and almost monodisperse assemblies
of cobalt or iron-oxide nano-particles and aim at measuring their
ferromagnetic resonance frequency and resonance field. It is then
desirable to have at least approximate but simple analytical formulas
to compare with the experimental results and to infer rough estimates
of the most relevant physical parameters, such as the elements size
and separation.

This paper has been organized as follows. In Section \ref{sec:General-formalism},
we define the system and its energy, focusing on the contribution
from the dipolar interactions. In Section \ref{sec:FMR-spectrum}
we present our general formalism in a matrix form and derive the final
expression for the frequency shift due to DI. Next, this formalism
is applied to a $2D$ array of nanomagnets and explicit expressions
are then given for the various contributions to the energy and for
the frequency shift, in particular in the case of thin disks. We also
discuss the contribution that stems from the size and shape of the
nano-elements, which adds up to the contribution that obtains within
the point-dipole approximation. In Section \ref{sec:Surface-effects}
we present the effective model for angular isolated nano-element and
discuss the surface contribution to the frequency shift. Section \ref{sec:Results-and-discussion}
shows some results of the comparison between the numerical and semi-analytical
calculations of the frequency shift. We also discuss the effect of
the array size on the difference in frequency shift between the results
of the two approaches. Finally, we discuss the competition between
DI and surface effects in two situations with a positive or negative
contribution from the latter. The paper ends with our conclusions
and two short appendices.\label{sec:General-formalism}

\subsection{Energy}

Here we define the systems targeted by this study and discuss the
various contributions to their energy, with a special focus on the
dipolar interactions. For simplicity, the discussion of the contribution
to the energy from the nano-element surface anisotropy is postponed
to Section \ref{sec:Surface-effects}.

Consider a monodisperse array of magnetic nano-elements each (of volume
$V$) carrying a magnetic moment ${\bf m}_{i}=m_{i}{\bf s}_{i},\,i=1,\ldots,\mathcal{{\cal N}}$
of magnitude $m_{i}=M_{s}V$ and direction ${\bf s}_{i}$, with $\vert{\bf s}_{i}\vert=1$,
$M_{s}$ being the saturation magnetization. The energy (in S.I. units)
of the magnetic moment ${\bf m}_{i}$ is given by 
\begin{equation}
E_{i}=E_{i}^{\left(0\right)}+E_{{\rm DI},i},\label{eq:TotalEnergy}
\end{equation}
where $E_{i}^{\left(0\right)}$ is the energy of the non-interacting
nano-elements that comprises the Zeeman and (effective) anisotropy
energies, and the second term $E_{{\rm DI}}$ is the DI contribution.
The total energy of the system is $E=\sum_{i=1}^{\mathcal{N}}E_{i}$.

For two magnetic nano-elements carrying macroscopic moments ${\bf m}_{i}$
and ${\bf m}_{j}$, located at two arbitrary sites $i$ and $j$,
the dipolar interaction reads (in SI units)

\begin{equation}
E_{{\rm DI},i,j}\equiv\left(\frac{\mu_{0}}{4\pi}\right){\bf m}_{i}\cdot{\cal D}_{ij}\cdot{\bf m}_{j}\label{eq:DI-PairEnergy}
\end{equation}
where ${\cal D}_{ij}$ is the corresponding tensor {[}see Eq. (\ref{eq:HorizDimerEnergy}){]}.
Summing over all pair-wise interactions, avoiding double counting,
yields the energy of a magnetic moment at site $i$ due to its interaction
with all other moments in the assembly with the corresponding energy
\begin{equation}
E_{{\rm DI},i}\equiv\left(\frac{\mu_{0}}{4\pi}\right)\sum_{j<i}{\bf m}_{i}\cdot{\cal D}_{ij}\cdot{\bf m}_{j}.\label{eq:DI-LocalEnergy}
\end{equation}

If one denotes by $R_{ij}$ the distance between the sites $i$ and
$j$ and use ${\bf m}_{i}=m_{i}{\bf s}_{i},{\bf m}_{j}=m_{j}{\bf s}_{j}$,
we see that $E_{{\rm DI},i}$ scales as $m^{2}/R^{3}$. Next, we may
introduce the distance $d$ as the nearest-neighbor inter-particle
separation, or the ``super-lattice'' parameter, and write \textcolor{black}{$R_{ij}=r_{ij}d$
where now $r_{ij}$ is a dimensionless parameter, which is calculated
as usual using only the integer indices used to locate a site on a
given lattice. More precisely, a site $i$ on discrete $2D$ lattice,
for instance, can be located using its coordinates $x_{i},y_{i}\in\mathbb{R}$
or by the corresponding integer indices $i_{x},i_{y}\in\mathbb{N}$.
Then, $R_{ij}=\sqrt{\left(x_{i}-x_{j}\right)^{2}+\left(y_{i}-y_{j}\right)^{2}}$
while $r_{ij}=\sqrt{\left(i_{x}-j_{x}\right)^{2}+\left(i_{y}-j_{y}\right)^{2}}=R_{ij}/d$.
Therefore, when dealing with DI on a super-lattice, it is quite natural
to introduce the parameter }

\textcolor{black}{
\begin{equation}
\lambda\equiv\left(\frac{\mu_{0}}{4\pi}\right)\frac{m^{2}}{d^{3}},\label{eq:DI-intensity}
\end{equation}
to characterize the strength of the DI in the system since the dipolar
energy in Eq. (\ref{eq:DI-PairEnergy}) scales with $\lambda$. }

In the next section, we will apply our formalism to specific situations
where the expressions of all contributions to the energy can be explicitly
written. 

In the present work, we develop a general formalism that can be applied
to an arbitrary system of interacting arrays of nano-magnets. However,
here it will be applied to the specific case of a two-dimensional
mono-disperse array of nano-magnets, with the main objective to derive
explicit expressions for the FMR frequency shift induced by the DI
and SE. These calculations are based on perturbation theory for a
dilute assembly but are valid for nano-magnets of arbitrary size and
shape, thus going beyond the simple point-dipole approximation (PDA).
On other hand, in Section \ref{sec:Surface-effects} we discuss in
detail surface effects that come into play when the size of the nano-elements
becomes small enough (with the number of surface atoms exceeding $50\%$).
In the framework of the effective model discussed earlier, we will
discuss how the expressions in the present section are extended to
include the contribution from surface anisotropy.

To summarize, our simplification only refer to:\emph{ i)} a collective
condition which assumes that the assembly is diluted. That is to say
the center-to-center distance between the nanoelements is much larger
than the linear dimension of the nanoelements. \emph{ii)} An intrinsic
condition requiring that the magnetic state of the nanoelements is
nearly saturated by the applied magnetic field. Now, the geometry
of the nanoelements themselves or that of the sample (\emph{i.e. }the
assembly thereof) are brought in by the dipolar tensor $\mathcal{D}$
and the distribution of the distances $r_{ij}$. 

\begin{figure}[H]
\begin{centering}
\includegraphics[width=0.8\columnwidth]{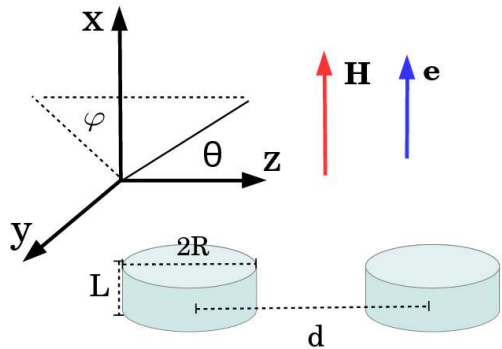} 
\par\end{centering}
\protect\caption{\foreignlanguage{american}{\label{fig:Dimer}A pair of magnetic nano-elements belonging to the
$2D$ array (in the $yz$ plane), of diameter $D=2R$, height $L$
and separation $d$. The standard system of spherical coordinates
$\left(\theta,\varphi\right)$ is also shown together with the setup
of the magnetic field $\bm{H}$ and \foreignlanguage{english}{(magneto-crystalline)}
anisotropy easy axes $\bm{e}$.}}
\end{figure}

Therefore, we consider a $2D$ array of magnetic nano-elements which
we assume to be lying in the $yz$ plane, for mathematical convenience.
The applied magnetic field and the (magneto-crystalline) uniaxial
anisotropy easy axes $\mathbf{e}_{i}$ are all directed along the
$x$ axis, \emph{i.e.} $\bm{H}=H\mathit{\mathbf{e}_{x}}$ and $\mathbf{e}_{i}=\mathit{\mathbf{e}_{x}}$
for all $i=1,\ldots,\mathcal{N}$. As shown in Fig. \ref{fig:Dimer},
we adopt the usual spherical coordinates for the magnetization orientation
\begin{eqnarray}
\mathbf{s}_{i} & = & \begin{cases}
s_{ix}=\sin\theta_{i}\cos\varphi_{i} & =s_{i}^{\perp}\cos\varphi_{i},\\
s_{iy}=\sin\theta_{i}\sin\varphi_{i} & =s_{i}^{\perp}\sin\varphi_{i},\\
s_{iz}=\cos\theta_{i} & =\cos\theta_{i},
\end{cases}\label{eq:new moment coordinates}
\end{eqnarray}
with $\left|\bm{s}_{i}\right|=1$.

In Eq. (\ref{eq:TotalEnergy}) the energy density $E_{i}^{\left(0\right)}$
of an isolated nano-element (ignoring its SE) is given by

\begin{equation}
E_{i}^{\left(0\right)}=-\mu_{0}M_{s}\bm{H}\cdot{\bf s}_{i}-K_{{\rm 2}}\left({\bf s}_{i}\cdot\bm{e}_{i}\right)^{2}+E_{{\rm demag}}\label{eq:En_FreeAssembly}
\end{equation}
where $K_{{\rm 2}}$ is the magneto-crystalline (uniaxial) anisotropy
constant, $\bm{e}_{i}$ the uniaxial anisotropy easy axis directed
along the $x$ axis (see Fig. \ref{fig:Dimer}). The term $E_{{\rm demag}}$
in Eq. (\ref{eq:En_FreeAssembly}) is the magnetostatic energy density
\begin{equation}
E_{{\rm demag}}=-\frac{\mu_{0}}{2}M_{s}\bm{H}_{{\rm d}}\cdot\mathbf{s}_{i}=\frac{\mu_{0}}{2}M_{s}^{2}\mathbf{s}_{i}\bm{N}\cdot\mathbf{s}_{i}\label{eq:Energy_demag}
\end{equation}
where $\bm{N}$ is the demagnetization tensor and $\bm{H_{\mathrm{d}}}$
the demagnetizing field. 

\textcolor{black}{A rigorous evaluation of the demagnetization tensor
for uniformly magnetized particles with cylindrical symmetry was provided
in Refs. {[}\onlinecite{tandonetalI04jmmm}{]} using elliptical integrals
{[}see Eq. (74) therein{]}. However, as already emphasized earlier,
our goal here is to derive an approximate analytical expression for
the FMR frequency shift due to DI. Now, FMR measurements are performed
under a DC magnetic field that is strong enough for saturating the
magnetic system, and this leads to a smoothing out of the spin non-colinearities
that usually occur in a magnetic system, especially when its aspect
ratio differs from unity. In addition, in our system setup, the external
magnetic field is applied in the direction of uniaxial anisotropy,
thus leading to a strong effective field along the cylinder axis.
In such a situation, the calculations of the demagnetizing field greatly
simplify, as is exemplified by Eqs. (26) and (27) of Ref. {[}\onlinecite{Caciagli_JMMM_2018}{]}.
Consequently, after averaging over the sample's length, the following
approximate expression for the demagnetization factors for a nano-element
with cylindrical symmetry about the $x$ axis (as is the case here)
is obtained \cite{Wysin_web}}

\begin{align*}
N_{x} & =\left(1+\delta\right)-\sqrt{1+\delta^{2}},\\
N_{y}= & N_{z}=\frac{1}{2}\left(\sqrt{1+\delta^{2}}-\delta\right),
\end{align*}
where $\delta\equiv R/L$, with $R$ being the radius of the cylinder
and $L$ its length (or thickness). In particular, for a very long
cylinder with $R\ll L$ ($\delta\ll1$), the longitudinal demagnetization
factor $N_{x}\rightarrow0$ while the transverse factors $N_{z}=N_{y}\rightarrow1/2$.
In the opposite limit, for a very thin disk, $R\gg L$ ($\delta\gg1$),
$N_{x}\rightarrow1$ and $N_{z}=N_{y}\rightarrow0$.

Therefore, using $\vert{\bf s}_{i}\vert=1$ the demagnetizing energy
density becomes (up to a constant) 
\begin{eqnarray}
E_{{\rm demag}} & = & \frac{\mu_{0}}{2}M_{s}^{2}\left(N_{x}-N_{z}\right)s_{i,x}^{2}.\label{eq:Energy_demag2}
\end{eqnarray}

The effective field $\bm{H}_{\mathrm{eff},i}=-\frac{1}{M_{s}}\delta E_{i}^{\left(0\right)}/\delta{\bf s}_{i}$,
normalized with respect to the anisotropy field 
\begin{equation}
\mu_{0}H_{K}=\frac{2K_{2}}{M_{s}},\label{eq:AnisotropyField}
\end{equation}
namely $\bm{H}_{\mathrm{eff},i}\longrightarrow\bm{h}_{\mathrm{eff},i}\equiv\bm{H}_{\mathrm{eff},i}/H_{K}$,
and upon dropping the index $i$ (for simplicity), reads 
\begin{align}
\bm{h}_{\mathrm{eff}} & =\left[h+ks_{x}+h_{d}s_{x}\right]\bm{e}_{x}\label{eq:heff}
\end{align}
where $h\equiv H/H_{K},h_{\mathrm{d}}\equiv-\mu_{0}M_{s}\left(N_{x}-N_{z}\right)/H_{K}$
and $k$ ($=0$ or $1$) is a label merely introduced for keeping
track of the contribution from magneto-crystalline anisotropy.

The angular frequency of an isolated nano-element is given by $\omega^{\left(0\right)}=\gamma H_{\mathrm{eff}}$,
with $\gamma\simeq1.76\times10^{11}\left({\rm T.s}\right)^{-1}$ being
the gyromagnetic ratio. Since, in the present setup, the minimum-energy
state of the nano-magnet corresponds to having its magnetic moment
along the $x$ axis, we have $\omega^{\left(0\right)}=\omega_{K}\left(h+k+h_{\mathrm{d}}\right)$,
where $\omega_{K}\equiv\gamma H_{K}$. Thus, for convenience we also
introduce the dimensionless angular frequency 
\begin{equation}
\varpi^{\left(0\right)}\equiv\frac{\omega^{\left(0\right)}}{\omega_{K}}=h+k+h_{\mathrm{d}}\label{eq:Frequency-free}
\end{equation}

In the sequel, we shall measure all frequencies in units of $\omega_{K}$,
\emph{i.e}, $\varpi\equiv\omega/\omega_{K}$.

\subsection{Dipolar interactions beyond the point-dipole approximation}

For a pair of magnetic nano-elements belonging to the $2D$ array,
as shown in Fig. \ref{fig:Dimer}, the DI interaction was obtained
in this horizontal configuration in Ref. \onlinecite{francoetal14jap},
see also Ref. \onlinecite{beleggiaetal04jmmm278}. It is given by

\begin{equation}
\mathcal{E}_{{\rm DI}}\equiv\frac{E_{{\rm DI}}}{2K_{{\rm 2}}V}=\frac{1}{2}\overset{\mathcal{N}}{\sum_{i=1}}\overset{\mathcal{N}}{\sum_{\begin{array}{c}
j=1\\
j\neq i
\end{array}}}\xi_{ij}\,\mathbf{s}_{i}\cdot\mathcal{D}_{ij}\mathbf{s}_{j}\label{eq:HorizDimerEnergy}
\end{equation}
where

\[
\mathcal{D}_{ij}=\frac{J_{ij}-(2+\Phi_{ij})\hat{\boldsymbol{r}}_{ij}\hat{\boldsymbol{r}}_{ij}}{r_{ij}^{3}}
\]
is the usual DI dyadic and $\hat{\boldsymbol{r}}_{ij}$ the unit vector
connecting the sites $i$ and $j$, \emph{i.e. }$\hat{\boldsymbol{r}}_{ij}=\boldsymbol{r}_{ij}/r_{ij}$.
$J_{ij}$ is the diagonal matrix 
\begin{eqnarray*}
J_{ij} & = & \left(\begin{array}{ccc}
\Phi_{ij} & 0 & 0\\
0 & 1 & 0\\
0 & 0 & 1
\end{array}\right)
\end{eqnarray*}
and $\Phi_{ij}$ is a function of the size and shape of the nano-elements
as well as their separation {[}see Fig. \ref{fig:Dimer} (right){]}.
It is defined by\cite{beleggiaetal04jmmm278,francoetal14jap} 

\begin{eqnarray}
\Phi_{ij}\left(\eta_{ij},\tau\right) & \equiv & \frac{\mathcal{J}_{d}^{\mathrm{h}}\left(\eta_{ij},\tau\right)}{\mathcal{I}_{d}^{\mathrm{h}}\left(\eta_{ij},\tau\right)}\label{eq:ShapeFunction}
\end{eqnarray}
with

\begin{eqnarray}
\mathcal{I}_{d}^{\mathrm{h}}\left(\eta,\tau\right) & = & 16\eta^{2}\tau\intop_{0}^{\infty}\frac{dq}{q^{2}}J_{1}^{2}\left(q\right)J_{1}\left(2\eta\tau q\right)\left[1-\frac{\left(1-e^{-2q\tau}\right)}{2q\tau}\right],\nonumber \\
\mathcal{J}_{d}^{\mathrm{h}}\left(\eta,\tau\right) & = & 16\eta^{3}\tau\intop_{0}^{\infty}\frac{dq}{q^{2}}J_{1}^{2}\left(q\right)J_{0}\left(2\eta\tau q\right)\left(1-e^{-2\tau q}\right).\label{eq:PhiIntegrals}
\end{eqnarray}
$J_{0}$ and $J_{1}$ are the well known Bessel functions. We have
also introduced the following geometrical or ``aspect-ratio'' parameters
\begin{equation}
\eta_{ij}\equiv\frac{r_{ij}d}{L},\tau\equiv\frac{L}{2R}.\label{eq:DistanceParams}
\end{equation}

Note that $r_{ij}d$ is the center-to-center distance between the
pair of nano-elements on sites $i,j$, with $d$ being the (super-lattice)
step of the array ($r_{ij}$ are dimensionless real numbers). Finally,
the DI coefficient $\xi_{ij}$, appearing in Eq. (\ref{eq:HorizDimerEnergy}),
is given by {[}see Eq. (\ref{eq:DI-intensity}){]} 
\begin{equation}
\xi_{ij}\equiv\frac{\lambda}{2K_{{\rm 2}}V}\mathcal{I}_{d}^{\mathrm{h}}\left(\eta_{ij},\tau\right).\label{eq:DI-Coefficient}
\end{equation}

Now, we discuss the equilibrium state of the system. In general this
is obtained by minimizing the total energy of the system with respect
to all degrees of freedom. In the present case, we would have to minimize
the energy (\ref{eq:TotalEnergy}), summed over the whole lattice,
with respect to the $2\mathcal{N}$ angles $\theta_{i},\varphi_{i},i=1,\ldots,\mathcal{N}$.
In general, it is well know to the numerical-computing community that
minimization of such a multi-variate function is a formidable task
that requires a lot of efforts and high computing powers. Apart from
the latter, one of the reasons is that there are no ``automatic''
algorithms for finding the absolute minimum of the function and for
each situation, one has to guide the solver through some prescribed
path(s). Since the interactions are pair-wise, one could also proceed
by obtaining the energy minima for a dimer and then sum over the lattice.
The equilibrium state of a dimer with DI in several configurations
of anisotropy and applied field were thoroughly studied in Ref. \onlinecite{francoetal14jap}
and the various extrema were found in a closed form. However, it is
clear that for interactions of arbitrary intensity the state of an
($N+1$)-body system does not necessarily include the state of the
$N$-body system. For this reason the states obtained in Ref. \onlinecite{francoetal14jap}
cannot just be extended to the array studied here by summing over
the lattice index. As such, and as stated earlier, we resort to an
analytical treatment based on a few simplifying assumptions. Accordingly,
we first assume that the DI are weak enough as to allow for such an
extension. In fact, in FMR measurements, the applied magnetic field
is usually strong enough as to saturate the magnetic state of the
system, and this leads to the nearly linear branch of the resonance
frequency as a function of the amplitude of the applied field. This
is the situation that we adopt here. Usually the DI in such a $2D$
array (oblate assembly) favor a net magnetic moment in the plane of
the array. However, the strong effective anisotropy and the (relatively)
strong magnetic field are parallel to each other and perpendicular
to the array's plane, and should then lead to a reorientation of all
magnetic moments towards their direction. For this reason, the setup
in Fig. \ref{fig:Dimer} leads to the equilibrium state: $\theta_{i}=\frac{\pi}{2},\varphi_{i}=0,i=1,\ldots,\mathcal{N}$.

The effect of dipolar interactions in ordered and disordered low-dimensional
nanoparticle assemblies have been studied by many authors \citep{Varon_dipolar_scirep2013}.
In dense assemblies, the DI lead to various local magnetic orders
that depend on the lattice structure \citep{luttis46pr,Varon_dipolar_scirep2013}.
Under some conditions, they may also induce long-range order leading
to the so-called super-ferromagnetic state \citep{lisnak13icfpm8}.
Obviously, the situation we consider here is quite different in that
the concentration we assume is not high enough as to lead to the onset
of assembly-wide collective states. In addition, as argued earlier
we assume that the DI do not modify the equilibrium state as determined
by a competition between the applied field and the anisotropy. However,
even such a weak intensity of DI would be important to the dynamics
of the assembly since then the energy barriers and thereby the relaxation
rates would be affected.

\section{\label{sec:FMR-spectrum}FMR spectrum : general formalism}

In this section, we present the general formalism we have developed
in order to derive approximate analytical expressions for the shift
of the FMR frequency induced by dipolar interactions.

\subsection{Landau-Lifshitz equation and FMR eigenvalue problem}

The time evolution of the magnetization orientation $\mathbf{s}_{i}$
is governed by the damped (norm-conserving) Landau-Lifshitz equation
(LLE) 
\begin{equation}
\frac{d\mathbf{s}_{i}}{dt_{r}}=-\mathbf{s}_{i}\times\bm{h}_{\mathrm{eff},i}+\alpha\,\mathbf{s}_{i}\times\left(\mathbf{s}_{i}\times\bm{h}_{\mathrm{eff},i}\right),\label{eq:MSPDampedLLE}
\end{equation}
where $\bm{h}_{\mathrm{eff},i}$ now comprises both the free and interacting
parts of the system's energy. $\alpha$ $\left(\lesssim1\right)$
the phenomenological damping parameter and $t_{r}$ is the (dimensionless)
time defined by $t_{r}=t/t_{s}$, where $t_{s}=\omega_{K}^{-1}$ is
the nanoelement's characteristic timescale.

In order to compute the spectrum of the system excitations we may
proceed by linearizing the LLE (\ref{eq:MSPDampedLLE}) about the
equilibrium state. Indeed, we assume that the equilibrium state, which
minimizes the system's energy, denoted by $\left\{ \mathbf{s}_{i}^{\left(0\right)}\right\} _{i=1,\ldots,\mathcal{N}}$,
has been determined with the help of some analytical or numerical
technique. Then, we may write $\bm{s}_{i}\simeq\mathbf{s}_{i}^{\left(0\right)}+\delta\bm{s}_{i}$
and transform the differential equation (\ref{eq:MSPDampedLLE}) into
the following equation \citep{BastardisEtal-jpcm2017} 
\begin{equation}
\frac{d\left(\delta\mathbf{s}_{i}\right)}{d\tau}=\sum\limits _{k=1}^{\mathcal{N}}\left[\mathcal{H}_{ik}\mathcal{I}\left(\alpha\right)\right]\delta\mathbf{s}_{k},\quad i=1,\ldots,\mathcal{N}\label{eq:LinearizedLLEMatrixForm}
\end{equation}
with the (pseudo-)Hessian 
\begin{eqnarray}
\mathcal{H}_{ik}\left[\mathcal{E}\right] & \equiv & \left(\begin{array}{ll}
\partial_{\theta_{k}\theta_{i}}^{2}\mathcal{E} & \frac{1}{\sin\theta_{i}}\partial_{\theta_{k}\varphi_{i}}^{2}\mathcal{E}\\
\\
\frac{1}{\sin\theta_{k}}\partial_{\varphi_{k}\theta_{i}}^{2}\mathcal{E} & \frac{1}{\sin\theta_{k}\sin\theta_{i}}\partial_{\varphi_{k}\varphi_{i}}^{2}\mathcal{E}
\end{array}\right)\label{eq:pseudo-Hessian}
\end{eqnarray}
whose matrix elements are second-order derivatives of the energy with
respect to the spherical angles of $\left(\theta_{i},\varphi_{i}\right)$
that determine the direction $\bm{s}_{i}$. All these matrix elements
are evaluated at the equilibrium state $\left\{ \mathbf{s}_{i}^{\left(0\right)}\right\} _{i=1,\ldots,\mathcal{N}}$.
The symbol $\partial_{\alpha_{k}\beta_{i}}^{2}$ stands for the second
derivative with respect to the angles $\alpha_{k},\beta_{i}$.

The matrix 
\[
\mathcal{I}\left(\alpha\right)\equiv\left(\begin{array}{cc}
\alpha & -1\\
1 & \alpha
\end{array}\right)
\]
stems from the double cross product in Eq. (\ref{eq:MSPDampedLLE}).

Next, one may seek solutions of Eq. (\ref{eq:LinearizedLLEMatrixForm})
in the form $\delta\mathbf{s}_{k}=\delta\mathbf{s}_{k}(0)\,e^{i\Omega\tau}$
leading to the following eigenvalue problem 
\begin{eqnarray}
\sum\limits _{k=1}^{\mathcal{N}}\left[\mathcal{H}_{ik}\mathcal{I}\left(\alpha\right)-i\Omega\mathds{1}\right]\delta\mathbf{s}_{k} & = & 0\label{eq:EVProblem}
\end{eqnarray}
whose set of roots $\left\{ \Omega_{n}\right\} _{1\leq n\leq\mathcal{N}}$
yields the system's eigen-frequencies, $f_{n}=-i\frac{\Omega_{n}}{2\pi}=\frac{\omega_{n}}{2\pi}$,
where $\omega_{n}$ is the real angular frequency (in rad/s). Here,
$\mathds{1}$ is the identity matrix with matrix elements $\mathds{1}{}_{ik}^{\mu\nu}=\delta_{ik}\delta^{\mu\nu}$.

In the general case of an array of nano-elements with arbitrary DI,
it is not possible to determine the system's exact equilibrium state
$\left\{ \mathbf{s}_{i}^{\left(0\right)}\right\} _{i=1,\ldots,\mathcal{N}}$
that minimizes the total energy of the system, including the (core
and surface) anisotropy, the DI and the applied field. In addition,
it is not an easy matter to solve the eigenvalue problem (\ref{eq:EVProblem})
in its full generality. Of course, these tasks can be numerically
accomplished to some extent. However, as stated earlier, our objective
here is to obtain an analytical expression for the FMR frequency.
In order to do so, we restrict ourselves to the case of dilute assemblies
of nanomagnets, and as such, we solve the eigenvalue problem (\ref{eq:EVProblem})
using perturbation theory that we present now.

\subsection{DI correction to FMR frequency : perturbation theory}

\textcolor{black}{In the following}\textcolor{blue}{{} }we only consider
the undamped case, \emph{i.e.} with $\alpha=0$ {[}see Eq. (\ref{eq:MSPDampedLLE}){]},
we present our formalism in the general case.

As we have seen, the excitation spectrum can be obtained by diagonalizing
the matrix $\mathbb{H}_{ik}\left(\alpha\right)\equiv\mathcal{H}_{ik}\mathcal{I}\left(\alpha\right)$
of matrix elements 
\[
\mathbb{H}_{ik}^{\mu\nu}\left(\alpha\right)=\sum_{\rho=\theta,\varphi}\mathcal{H}_{ik}^{\mu\rho}\mathcal{I}^{\rho\nu}\left(\alpha\right).
\]

A word is in order regarding the various indices. The problem being
studied here is an array of magnetic moments located at the nodes
of a super-lattice. Hence, there are two kinds of indices. The first
one, using the Greek letters $\mu,\nu,\rho$, refers to the components
of a magnetic moment in the system of spherical coordinates and thus
assumes the values $\theta,\varphi$. The second index, using the
Roman letters $i,j,k$, refers to the lattice site and assumes the
values $1,\ldots,\mathcal{N}$. Therefore, the full ``phase space''
is a direct product of the two sub-spaces corresponding to the two
kinds of variables. Likewise, the matrices involved in these calculations
are tensor products of the corresponding sub-matrices.

If we focus on dilute assemblies with relatively weak DI we can write
the total energy as the sum of the energy of the non-interacting assembly
and the interaction contribution, \emph{i.e.} $\mathcal{E}=\mathcal{E}^{\left(0\right)}+\mathcal{E}_{{\rm DI}}$,
with $\mathcal{E}\equiv E/\left(2K_{{\rm 2}}V\right)$ and similarly
for each contribution. Then, the pseudo-Hessian $\mathcal{H}_{ik}(\mathcal{E}_{i})$
can also be correspondingly split as follows 
\begin{equation}
\mathbb{H}_{ik}^{\mu\nu}\left(\alpha\right)=\mathcal{F}_{ik}^{\mu\nu}+\Xi_{ik}^{\mu\nu}\label{eq:PerturbationTheory}
\end{equation}
where $\mathcal{F}_{ik}$ is the contribution in the absence of interactions
given by the same matrix as in Eq. (\ref{eq:pseudo-Hessian}), upon
substituting $\mathcal{E}^{\left(0\right)}$ for $\mathcal{E}$, and
multiplied by $\mathcal{I}\left(\alpha\right)$. Thus, $\mathcal{F}=\mathcal{H}\left[\mathcal{E}^{\left(0\right)}\right]\mathcal{I}\left(\alpha\right)$.
Similarly, $\Xi_{ik}$ is the DI contribution given by the matrix
in Eq. (\ref{eq:pseudo-Hessian}), with substitution of $\mathcal{E}_{{\rm DI}}$
for $\mathcal{E}$, multiplied by $\mathcal{I}\left(\alpha\right)$,
\emph{i.e.} $\Xi=\mathcal{H}\left[\mathcal{E}_{{\rm DI}}\right]\mathcal{I}\left(\alpha\right)$.
For later use, we introduce the two matrices $F\equiv\mathcal{H}\left[\mathcal{E}^{\left(0\right)}\right],\Theta\equiv\mathcal{H}\left[\mathcal{E}_{{\rm DI}}\right]$.

It is understood that wherever they appear all matrix elements have
to be evaluated at the equilibrium state $\left\{ \mathbf{s}_{i}^{\left(0\right)}\right\} _{i=1,\ldots,\mathcal{N}}$
with $\mathbf{s}_{i}^{\left(0\right)}=\left(1,0,0\right)$. In the
situation of relatively weak coupling considered here, we make the
further assumption that the equilibrium state is not altered by the
dipolar interactions. More precisely, we assume that the main equilibrium
of the system is setup by the competition between the strong (effective)
anisotropy and the external DC magnetic field. Of course, the DI of
arbitrary strength would change both the energy minima and saddle
points of the system, and thereby significantly change its dynamics.
Here we restrict ourselves to the situation where the DI only contribute
through the second term in Eq. (\ref{eq:PerturbationTheory}), which
is regarded as a correction to the first term. This assumption is
experimentally relevant for dilute assemblies. For instance, it has
been demonstrated\citep{Yang_etal_APL2009} that inter-particle interactions
in assemblies of core-shell (Fe$_{3}$O$_{4}$/SiO$_{2}$) nanoparticles
can be tuned by modifying the thickness of the shell.

Therefore, in spin components the LLE (\ref{eq:LinearizedLLEMatrixForm})
reads 
\begin{eqnarray}
\frac{d\left(\delta s_{i}^{\mu}\right)}{dt} & = & \sum_{\nu=\theta,\varphi}\sum\limits _{k=1}^{\mathcal{N}}\left[\mathcal{F}\left(\mathds{1}+\mathcal{F}^{-1}\Xi\right)\right]_{ik}^{\mu\nu}\delta s_{k}^{\nu}.\label{eq:linLLETensorForm}
\end{eqnarray}

Regarding the various indices discussed above, the matrix $\mathcal{F}$
of the non-interacting array can be written as $\mathcal{F}=\mathcal{F}_{2\times2}\otimes\mathds{1}_{\mathcal{N}\times\mathcal{N}}$,
or in components $\mathcal{F}_{ik}^{\mu\nu}=\left(\mathcal{F}_{2\times2}^{\mu\nu}\right)_{i}\delta_{ik}=\delta_{ik}\mathcal{H}\left[\mathcal{E}^{\left(0\right)}\right]_{ii}^{\mu\nu}\mathcal{I}\left(\alpha\right).$
Using the matrix $F$ introduced earlier, the matrix $\mathcal{F}$
explicitly reads 
\[
\mathcal{F}_{ik}=\delta_{ik}\left(\begin{array}{cc}
F_{ii}^{\theta\varphi} & -F_{ii}^{\theta\theta}\\
F_{ii}^{\varphi\varphi} & -F_{ii}^{\varphi\theta}
\end{array}\right)\otimes\mathds{1}.
\]

Note that the $2\times2$ matrix above has two eigenvalues $\pm i\varpi_{i}^{\left(0\right)}$,
where $\varpi_{i}^{\left(0\right)}$ is the (normalized) resonance
frequency of the magnetic moment at site $i$ in the non-interacting
case. In fact, for the mono-disperse assemblies considered here all
these frequencies are identical, \emph{i.e.} $\varpi_{i}^{\left(0\right)}\equiv\varpi^{\left(0\right)}$,
defined in Eq. (\ref{eq:Frequency-free}). Hence, $\det\mathcal{F}=\prod_{i}^{\mathcal{N}}\left(\varpi_{i}^{\left(0\right)}\right)^{2}=\left(\varpi^{\left(0\right)}\right)^{2\mathcal{N}}$.

On the other hand, the matrix $\Xi$ introduced above and which contains
the DI contribution can also be written explicitly to some limit.
Again, using the matrix $\Theta$ introduced earlier, the $2\times2$
diagonal block of the matrix $\Xi=\mathcal{H}\left[\mathcal{E}_{{\rm DI}}\right]\mathcal{I}\left(\alpha\right)$
reads 
\begin{equation}
\left(\begin{array}{cc}
\Theta_{ii}^{\theta\varphi} & -\Theta_{ii}^{\theta\theta}\\
\Theta_{ii}^{\varphi\varphi} & -\Theta_{ii}^{\varphi\theta}
\end{array}\right),\quad i=1,2,\ldots,\mathcal{N}.\label{eq:XiElementsDiag}
\end{equation}

One should note that these DI matrix elements with identical lattice
sites are not equal to zero even if they correspond to pair-wise interactions.
Indeed, in the most general situation, the second derivatives of the
energy are given by

\begin{widetext}

\begin{eqnarray}
\partial_{\theta_{i}\theta_{k}}^{2}\mathcal{E} & = & \delta_{ik}\left[\mathbf{s}_{i}\cdot-\mathbf{e}_{\theta_{i}}\cdot\left(\mathbf{e}_{\theta_{i}}\cdot\mathbf{\nabla}_{i}\right)\right]\bm{h}_{\mathrm{eff},i}-\left(1-\delta_{ik}\right)\mathbf{e}_{\theta_{i}}\cdot\left[\mathbf{e}_{\theta_{k}}\cdot\mathbf{\nabla}_{k}\right]\bm{h}_{\mathrm{eff},i},\nonumber \\
\partial_{\varphi_{k}\varphi_{i}}^{2}\mathcal{E} & = & \delta_{ik}\sin\theta_{i}\left[\left(\sin\theta_{i}\,\mathbf{s}_{i}+\cos\theta_{i}\,\mathbf{e}_{\theta_{i}}\right)-\sin\theta_{i}\,\mathbf{e}_{\varphi_{i}}\cdot\left(\mathbf{e}_{\varphi_{i}}\cdot\mathbf{\nabla}_{i}\right)\right]\bm{h}_{\mathrm{eff},i}-\left(1-\delta_{ik}\right)\sin\theta_{i}\sin\theta_{k}\,\mathbf{e}_{\varphi_{i}}\cdot\left[\mathbf{e}_{\varphi_{k}}\cdot\mathbf{\nabla}_{k}\right]\bm{h}_{\mathrm{eff},i},\nonumber \\
\partial_{\theta_{k}\varphi_{i}}^{2}\mathcal{E} & = & -\delta_{ik}\left[\cos\theta_{i}\,\mathbf{e}_{\varphi_{i}}\cdot+\sin\theta_{i}\,\mathbf{e}_{\varphi_{i}}\cdot\left(\mathbf{e}_{\theta_{i}}\cdot\mathbf{\nabla}_{i}\right)\right]\bm{h}_{\mathrm{eff},i}-\left(1-\delta_{ik}\right)\sin\theta_{i}\,\mathbf{e}_{\varphi_{i}}\cdot\left[\mathbf{e}_{\theta_{k}}\cdot\mathbf{\nabla}_{k}\right]\bm{h}_{\mathrm{eff},i},\nonumber \\
\partial_{\varphi_{k}\theta_{i}}^{2}\mathcal{E} & = & -\delta_{ik}\left[\cos\theta_{i}\,\mathbf{e}_{\varphi_{i}}\cdot+\sin\theta_{i}\,\mathbf{e}_{\theta_{i}}\cdot\,\left[\mathbf{e}_{\varphi_{i}}\cdot\mathbf{\nabla}_{i}\right]\right]\bm{h}_{\mathrm{eff},i}-\left(1-\delta_{ik}\right)\sin\theta_{k}\,\mathbf{e}_{\theta_{i}}\cdot\left(\mathbf{e}_{\varphi_{k}}\cdot\mathbf{\nabla}_{k}\right)\bm{h}_{\mathrm{eff},i}.\label{eq:Energy2Derivatives}
\end{eqnarray}

\end{widetext}

Explicit expressions for the DI energy only are given in Appendix
\ref{sec:The-pseudo-Hessian-matrix}. Note then that because of the
first term in each line of Eq. (\ref{eq:Energy2Derivatives}), the
second derivatives $\Theta_{ii}^{\mu\nu}$ do not vanish even for
the DI contribution, and using (\ref{eq:pseudo-Hessian}, \ref{eq:SecondDerivsDI})
we do see that $\Theta_{ii}^{\mu\nu}\neq0$. However, we stress that
the DI contribution to $\bm{h}_{\mathrm{eff},i}$ contains a sum over
the whole lattice except (for the site $i$) and thus the DI coefficient
entering $\bm{h}_{\mathrm{eff},i}$ involves a sum over $j$ with
$j\neq i$. Obviously, the matrix $\Xi$ has also nonzero off-diagonal
blocks which are of the same form as in (\ref{eq:XiElementsDiag})
but with distinct indices $i,k=1,2,\ldots,\mathcal{N}$, \emph{i.e.}
$i\neq k$.

Now, we introduce the new tensor $\Lambda\equiv\mathcal{F}\left(\mathds{1}+\mathcal{F}^{-1}\Xi\right)$
\begin{equation}
\det\Lambda=\det\mathcal{F}\times\det\left(\mathds{1}+\mathcal{F}^{-1}\Xi\right).\label{eq:DetLambda}
\end{equation}
and set to compute its determinant. Similarly to $\varpi_{n}^{\left(0\right)}$
(the eigenvalues of $\mathcal{F}$) we introduce the eigenvalues $\varpi_{n}$
as the resonance frequencies with the index $n$ running through all
the $2\mathcal{N}$ (collective) modes of the interacting system.
This leads to $\det\Lambda=\prod_{n=1}^{\mathcal{N}}\varpi_{n}^{2}$.
Then, let us examine the last determinant in Eq. (\ref{eq:DetLambda}).
The product $\mathcal{F}^{-1}\Xi$ scales with the ratio $\lambda/H$,
\emph{i.e. }the ratio of the DI intensity $\lambda=\left(\frac{\mu_{0}}{4\pi}\right)m^{2}/d^{3}$
to the static magnetic field $H$. This ratio is obviously small for
a dilute assembly, especially for standard FMR measurements where
the DC field is usually taken strong enough to saturate the sample
(usually between $0.3\,\mathrm{T}\;\mathrm{and}\;1\,{\rm T}$). Hence,
it is justified to make an expansion with respect to $\mathcal{F}^{-1}\Xi$.
For this, we apply the logarithm and use the expansion $\log\left(1+x\right)\simeq x$
(for operators) together with the identity $\log\det A={\rm Tr}\log A$.
Doing so, we obtain 
\begin{align}
\sum_{i=1}^{\mathcal{N}}\log\varpi_{i}^{2} & \simeq\sum_{i=1}^{\mathcal{N}}\log\left(\varpi_{i}^{\left(0\right)}\right)^{2}+{\rm Tr}\left[\mathcal{F}^{-1}\Xi\right]\nonumber \\
 & =2\mathcal{N}\log\varpi^{\left(0\right)}+{\rm Tr}\left[\mathcal{F}^{-1}\Xi\right].\label{eq:LogSumBase}
\end{align}

In order to compute the trace above, we only need to collect the (block)
diagonals of the matrix $\mathcal{F}^{-1}\Xi$ whose first block is
as follows (showing only the diagonal elements)

\[
\begin{array}{l}
\frac{1}{\left(\varpi^{\left(0\right)}\right)^{2}}\left(\begin{array}{cc}
-F_{ii}^{\varphi\theta} & F_{ii}^{\theta\theta}\\
-F_{ii}^{\varphi\varphi} & F_{ii}^{\theta\varphi}
\end{array}\right)\left(\begin{array}{cc}
\Theta_{ii}^{\theta\varphi} & -\Theta_{ii}^{\theta\theta}\\
\Theta_{ii}^{\varphi\varphi} & -\Theta_{ii}^{\varphi\theta}
\end{array}\right)\\
\\
=\frac{1}{\left(\varpi^{\left(0\right)}\right)^{2}}\left(\begin{array}{cc}
F_{ii}^{\theta\theta}\Theta_{ii}^{\varphi\varphi}-F_{ii}^{\varphi\theta}\Theta_{ii}^{\theta\varphi} & *\\*
* & F_{ii}^{\varphi\varphi}\Theta_{ii}^{\theta\theta}-F_{ii}^{\theta\varphi}\Theta_{ii}^{\varphi\theta}
\end{array}\right)
\end{array}
\]
and thereby we obtain

\[
\begin{array}{lll}
{\rm Tr}\left[\mathcal{F}^{-1}\Xi\right] & = & \frac{1}{\left(\varpi^{\left(0\right)}\right)^{2}}{\displaystyle \sum_{i=1}^{\mathcal{N}}\left[F_{ii}^{\theta\theta}\Theta_{ii}^{\varphi\varphi}\right.}\\
\\
 &  & \left.-\left(F_{ii}^{\varphi\theta}\Theta_{ii}^{\theta\varphi}+F_{ii}^{\theta\varphi}\Theta_{ii}^{\varphi\theta}\right)+F_{ii}^{\varphi\varphi}\Theta_{ii}^{\theta\theta}\right].
\end{array}
\]
Thus, Eq. (\ref{eq:LogSumBase}) becomes

\[
\begin{array}{lll}
\frac{1}{\mathcal{N}}{\displaystyle \sum_{i=1}^{\mathcal{N}}}\log\varpi_{i} & = & \log\varpi^{\left(0\right)}+\frac{1}{2\mathcal{N}\left(\varpi^{\left(0\right)}\right)^{2}}{\displaystyle \sum_{i=1}^{\mathcal{N}}}\left[F_{ii}^{\theta\theta}\Theta_{ii}^{\varphi\varphi}\right.\\
\\
 &  & \left.-\left(F_{ii}^{\varphi\theta}\Theta_{ii}^{\theta\varphi}+F_{ii}^{\theta\varphi}\Theta_{ii}^{\varphi\theta}\right)+F_{ii}^{\varphi\varphi}\Theta_{ii}^{\theta\theta}\right].
\end{array}
\]
Next, it is quite reasonable to drop the sum on the left-hand side
of the equation above as long as one considers nano-element arrays
which are large enough and spatially isotropic. Then, upon expanding
with respect to the small parameter $\varpi_{i}/\varpi^{\left(0\right)}\lesssim1$,
we obtain the final expression for the DI-induced frequency shift
$\Delta\varpi_{{\rm {\rm DI}}}\equiv\varpi-\varpi^{\left(0\right)}$

\begin{equation}
\begin{array}{lll}
\Delta\varpi_{{\rm {\rm DI}}} & \simeq & \frac{1}{2\varpi^{\left(0\right)}}\frac{1}{\mathcal{N}}{\displaystyle \sum_{i=1}^{\mathcal{N}}\left[F_{ii}^{\theta\theta}\Theta_{ii}^{\varphi\varphi}\right.}\\
\\
 &  & \left.-\left(F_{ii}^{\varphi\theta}\Theta_{ii}^{\theta\varphi}+F_{ii}^{\theta\varphi}\Theta_{ii}^{\varphi\theta}\right)+F_{ii}^{\varphi\varphi}\Theta_{ii}^{\theta\theta}\right].
\end{array}\label{eq:DICorrection}
\end{equation}
We recall here again that the various matrix elements appearing above
are second derivatives of the energy with respect to the system coordinates,
evaluated at the equilibrium state $\left\{ \mathbf{s}^{\left(0\right)}\right\} $,
with $\mathbf{s}^{\left(0\right)}=\left(1,0,0\right)$. In some particular
situations of anisotropy and field setup, the matrix $\Xi$ can be
explicitly computed, thus directly rendering the correction to the
FMR frequency. Accordingly, in the next section we give explicit results
for the specific case of nano-elements with effective uniaxial anisotropy
along the field direction.

\subsection{\label{sec:FMR-2DArray}FMR frequency of a $2D$ array of nanomagnets}

\subsubsection{DI-induced frequency shift}

Now, we come to the evaluation of the matrix elements appearing in
Eq. (\ref{eq:DICorrection}).

For the non-interacting case we have

\[
F_{ii}=\left(\begin{array}{ll}
\partial_{\theta_{i}}^{2}\mathcal{E}_{i}^{\left(0\right)} & \frac{1}{\sin\theta_{i}}\partial_{\theta_{i}\varphi_{i}}^{2}\mathcal{E}_{i}^{\left(0\right)}\\
\\
\frac{1}{\sin\theta_{i}}\partial_{\varphi_{i}\theta_{i}}^{2}\mathcal{E}_{i}^{\left(0\right)} & \frac{1}{\sin^{2}\theta_{i}}\partial_{\varphi_{i}}^{2}\mathcal{E}_{i}^{\left(0\right)}
\end{array}\right)=\left(\begin{array}{ll}
\varpi^{\left(0\right)} & 0\\
0 & \varpi^{\left(0\right)}
\end{array}\right).
\]

For the DI contribution, both derivatives $\partial_{\theta_{i}}^{2}\mathcal{E}_{{\rm DI}}$
and $\partial_{\varphi_{i}}^{2}\mathcal{E}_{{\rm DI}}$ survive in
Eq. (\ref{eq:DICorrection}) when evaluated at the equilibrium state
$\left(\theta_{i}=\pi/2,\varphi_{i}=0\right)$, whereas the cross
derivatives vanish. Consequently, we obtain 
\[
\Theta_{ii}^{\theta\theta}=\Theta_{ii}^{\varphi\varphi}=-\left(\frac{\lambda}{2K_{{\rm 2}}V}\right)\overset{\mathcal{N}}{\sum_{\begin{array}{c}
j=1\\
j\neq i
\end{array}}}\frac{\mathcal{I}_{d}^{\mathrm{h}}\left(\eta_{ij},\tau\right)}{r_{ij}^{3}}\,\Phi_{ij}.
\]
Next, using (\ref{eq:ShapeFunction}) and introducing the geometrical
factor $\kappa$ as the ratio of the inter-element separation $d$
to their diameter $D=2R$, \emph{i.e.} $\kappa=d/D$ {[}see Eq. (\ref{eq:DI-intensity}){]},
we may rewrite the result above as follows 
\[
\Theta_{ii}^{\theta\theta}=\Theta_{ii}^{\varphi\varphi}=-\frac{A}{\kappa^{3}}\overset{\mathcal{N}}{\sum_{\begin{array}{c}
j=1\\
j\neq i
\end{array}}}\frac{\mathcal{J}_{d}^{\mathrm{h}}\left(\eta_{ij},\tau\right)}{r_{ij}^{3}}
\]
where we have introduced the material-dependent constant 
\[
A\equiv\left(\frac{\mu_{0}}{4\pi}\right)\frac{m^{2}/D^{3}}{2K_{{\rm 2}}V}=\left(\frac{\mu_{0}}{8\pi}\right)\frac{M_{s}^{2}V}{K_{{\rm 2}}D^{3}}.
\]

Then, substituting this result in Eq. (\ref{eq:DICorrection}) we
arrive at the explicit DI correction to the FMR (dimensionless) angular
frequency $\varpi$ of the array of nano-magnets 
\begin{equation}
\Delta\varpi_{{\rm {\rm DI}}}\simeq-\frac{A}{\kappa^{3}}\times\frac{1}{\mathcal{N}}\sum_{i=1}^{\mathcal{N}}\overset{\mathcal{N}}{\sum_{\begin{array}{c}
j=1\\
j\neq i
\end{array}}}\frac{\mathcal{J}_{d}^{\mathrm{h}}\left(\eta_{ij},\tau\right)}{r_{ij}^{3}}.\label{eq:DICorrection-v2}
\end{equation}

\subsubsection{$2D$ array of nano-disks}

The DI correction to the FMR frequency given in Eq. (\ref{eq:DICorrection-v2})
is an implicit expression that depends on various parameters pertaining
both to the nano-elements themselves (size, shape, energy) and to
the assembly (spatial arrangement and shape). In particular, the nano-elements
separation $d$ enters this expression via the integral $\mathcal{J}_{d}^{\mathrm{h}}\left(\eta_{ij},\tau\right)$
and the parameter $\kappa$. In order to derive an explicit (analytical)
expression for the frequency shift in terms of the nano-elements separation
$d$ (or the parameter $\kappa$), one has to (numerically) compute
the integrals $\mathcal{I}_{d}^{\mathrm{h}}\left(\eta,\tau\right)$
and $\mathcal{J}_{d}^{\mathrm{h}}\left(\eta,\tau\right)$. However,
this can also be analytically done in some limiting cases of the parameters
$\eta$ and $\tau$, namely $\tau\ll1$ for thin disks (or platelets)
or $\tau\gg1$ for long cylinders (or wires). By way of illustration,
in the present work we perform these calculations in the case of thin
disks.

For thin disks ($\tau\ll1$), the integrands in Eq.~(\ref{eq:PhiIntegrals})
decay to zero for\textcolor{blue}{{} }$q\gtrsim3$\textcolor{blue}{.}
Hence, we can expand the exponential in these integrals up to the
first order and then expand the integrals in powers of $1/\kappa$
(for $\kappa>1$). This yields

\begin{eqnarray}
\mathcal{J}_{d}^{\mathrm{h}}\left(\kappa_{ij},\tau\right) & \simeq & 1+\frac{9}{16\kappa^{2}}\times\frac{1}{r_{ij}^{2}},\nonumber \\
\mathcal{I}_{d}^{\mathrm{h}}\left(\kappa_{ij},\tau\right) & \simeq & 1+\frac{3}{16\kappa^{2}}\times\frac{1}{r_{ij}^{2}},\label{eq:JIDisks}
\end{eqnarray}
and

\begin{equation}
\Phi\left(\kappa_{ij},\tau\right)=\frac{\mathcal{J}_{d}^{\mathrm{h}}\left(\kappa_{ij},\tau\right)}{\mathcal{I}_{d}^{\mathrm{h}}\left(\kappa_{ij},\tau\right)}\simeq1+\frac{3}{8\kappa^{2}}\times\frac{1}{r_{ij}^{2}}.\label{eq:PhiDisks}
\end{equation}

To be specific, we consider the FeV disks of Ref. \onlinecite{pigeauetal12prl}
with $D=600\,{\rm nm},L=26.7\,{\rm nm}$ and a center-to-center separation
$d=1600\,{\rm nm}$, we have $\tau=1/\left(2\delta\right)=L/D=0.0445$,
$\eta=d/L\simeq60$ and thereby $\kappa=\eta\tau=d/D=2.667$. Therefore,
the condition for the validity of the results above, \emph{i.e.} $\kappa>1$
is satisfied even in the most unfavorable case.

Consequently, we obtain the frequency shift 
\begin{align}
\Delta\varpi_{{\rm {\rm DI}}} & \simeq-\frac{A}{\kappa^{3}}\left[\mathcal{C}_{3}+\frac{9}{16\kappa^{2}}\mathcal{C}_{5}\right]\label{eq:DICorrection-Disks}
\end{align}
where we have introduced the lattice sum 
\[
\mathcal{C}_{n}\equiv\frac{1}{\mathcal{N}}\sum_{i=1}^{\mathcal{N}}\overset{\mathcal{N}}{\sum_{\begin{array}{c}
j=1\\
j\neq i
\end{array}}}\frac{1}{r_{ij}^{n}}.
\]

We thus have to evaluate two lattice sums, the well-known one \citep{jongar01prb}
$\mathcal{C}_{3}\equiv\sum_{i}^{\mathcal{N}}\sum_{k,k\neq i}^{\mathcal{N}}\left[1/\left(\mathcal{N}r_{ik}^{3}\right)\right]$,
which is equal to\textcolor{red}{{} }\textcolor{black}{$\mathcal{C}_{3}\simeq$}
9 and the other $\mathcal{C}_{5}\equiv\sum_{i}^{\mathcal{N}}\sum_{k,k\neq i}^{\mathcal{N}}\left[1/\left(\mathcal{N}r_{ik}^{5}\right)\right]$,
equal to $\mathcal{C}_{5}\simeq5.1,$ both in the thermodynamic limit.

\begin{figure}
\begin{centering}
\includegraphics[width=0.95\columnwidth]{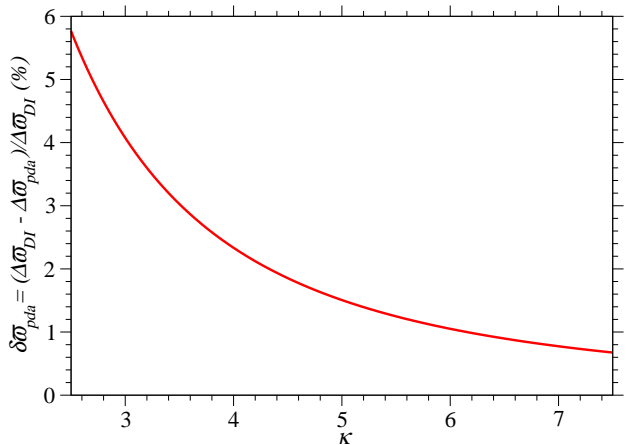}
\par\end{centering}
\caption{\label{fig:PDA_vs_DI}Relative variation of the frequency shift between
PDA and cylindrical nanomagnets $\delta\varpi_{{\rm pda}}=\frac{\left|\Delta\varpi_{{\rm DI}}-\Delta\varpi_{{\rm pda}}\right|}{\Delta\varpi_{{\rm DI}}}$
as a function of the distance $\kappa$. }

\end{figure}

We may then rewrite Eq. (\ref{eq:DICorrection-Disks}) as follows
(assuming $\mathcal{C}_{3}\neq0$, \emph{i.e.} excluding spheres and
cubes)

\begin{equation}
\Delta\varpi_{{\rm {\rm DI}}}\simeq\Delta\varpi_{{\rm pda}}\left(1+\frac{9}{16\kappa^{2}}\frac{\mathcal{C}_{5}}{\mathcal{C}_{3}}\right)\label{eq:DICorrection-Disks-2}
\end{equation}
where we have singled out the contribution $\Delta\varpi_{{\rm pda}}\equiv-\left(A/\kappa^{3}\right)\mathcal{C}_{3}$
that obtains within the PDA. As such, we see more explicitly the correction
to the FMR frequency due to the size and shape of the nanomagnets.
Both contributions in Eq. (\ref{eq:DICorrection-Disks-2}) are in
the form of a dipolar-like term multiplied by a lattice sum. While
the PDA term $\Delta\varpi_{{\rm pda}}$ scales with the nano-elements
separation $d$ as $1/d^{3}$, the term that stems from size and shape
effects scales with $d$ as $1/d^{5}$. The power $5$ here arises
from the $3$-dimensional space coordinates of the individual nano-particles,
plus the $2$ space dimensions arising from the shape of the disks,
for which the thickness is ignored (in the current thin-disk approximation).
Likewise, the expansions of the shape integrals (\ref{eq:JIDisks})
and (\ref{eq:PhiDisks}) also exhibit a point-dipole contribution
together with a $2$-dimensional shape correction that scales as $1/\kappa^{2}$.
\textcolor{black}{In Fig. \ref{fig:PDA_vs_DI} we plot the relative
difference between $\Delta\varpi_{{\rm DI}}$ and $\Delta\varpi_{{\rm pda}}$,
namely $\delta\varpi_{{\rm pda}}=\frac{\left|\Delta\varpi_{{\rm DI}}-\Delta\varpi_{{\rm pda}}\right|}{\Delta\varpi_{{\rm DI}}}$.
The results confirm that, for not-too-dense assemblies, }\textcolor{black}{\emph{i.e.}}\textcolor{black}{{}
for $2.5\lesssim\kappa\lesssim3$, there is a variation ($\delta\varpi_{{\rm pda}}\simeq5\%$)
of the frequency shift due to the fact that the nanomagnets are not
simple point dipoles. This variation should be accessible to experiments.
Obviously, for very dilute assemblies ($\kappa\gtrsim7$) the PDA
provides a correct description of the physics up to an error less
than 1\%. }

\section{\label{sec:Surface-effects}Surface effects}

In this section we discuss the impact of surface effects on the results
obtained above. In an assembly of nanomagnets the intrinsic features
of the latter, such as surface anisotropy (SA), are generally smoothed
out by the distributions of size and (easy axis) orientation. However,
in some situations, \emph{e.g.} of monodisperse assemblies with oriented
anisotropy, as is considered here, SE may lead to a non negligible
contribution to the magnetic properties of the nano-elements, especially
FMR frequency. Many examples of such assemblies have been fabricated
by several experimental groups around the world, see the already cited
works in the introduction as well Refs. \onlinecite{shendruketal07nanotech,
issaetal13imsc, khusrhidetal14jap, lavoratoetal15jpcc}. Surface effects
are local effects whose study requires recourse to an atomic approach
that accounts for the local atomic environment. However, from the
computational point of view, taking account of such effects in an
interacting assembly leads to tremendous difficulties which cannot
be efficiently dealt with even with the help of optimized numerical
approaches. Nonetheless, in the limiting case of not-too-strong surface
effects, inasmuch as the spin configuration inside of the nanomagnet
can be regarded as quasi-collinear, the static and dynamic properties
of the nanomagnet may be recovered with the help of an effective macroscopic
model for the net magnetic moment of the nanomagnet. More precisely,
it has been shown that a many-spin nanomagnet of a given lattice structure
and energy parameters (on-site core and surface anisotropy, local
exchange interactions) may be modeled by a macroscopic magnetic moment
${\bf m}$ evolving in an effective potential\cite{garkac03prl}.
The latter is, in principle an infinite polynomial in the components
of ${\bf m}$, but whose leading terms are of two types, one is a
quadratic and the other a quartic contribution with coefficients $K_{2}$
and $K_{4}$ that strongly depends on the microscopic parameters,
as well as on the shape and size of the nanomagnet. Here, we would
like to emphasize in passing the fact that the quartic term is a pure
surface contribution, that appears even in the absence of core anisotropy
{[}see Ref. \onlinecite{garkac03prl, kachkachi07j3m}{]} and which
may renormalize the cubic anisotropy of the (underlying) magnetic
material the nanomagnet is made of. However, there remains the question
as to how one can distinguish this surface-induced fourth-order contribution
from the (usually weak) cubic anisotropy found in magnetic materials.
At least for thin disks where the effective anisotropy is mostly of
(boundary) surface origin, this quartic contribution may become dominant.
An example of this situation was provided by cobalt nano-dots with
enhanced edge magnetic anisotropy \citep{rohartetal07prb}.

In the present work, we assume that the uniaxial anisotropy in Eq.
(\ref{eq:En_FreeAssembly}), with coefficient $K_{2}$, is an effective
anisotropy that already includes the (small) renormalization effect
from surface anisotropy. On the other hand, the strongest contribution
induced by surface effects is given by 
\begin{equation}
E_{i}^{\left({\rm SE}\right)}=\frac{1}{2}K_{4}\sum_{\alpha=x,y,z}s_{i,\alpha}^{4}.\label{eq:SE}
\end{equation}
where $K_{4}$ is a constant that scales with the square of the surface
anisotropy constant\citep{garkac03prl}. $K_{4}$ may be positive
or negative, depending on the underlying magnetic material\citep{yanesetal07prb}.
In the sequel, we will use the more relevant parameter $\zeta\equiv K_{4}/K_{2}$.
Consequently, adding this contribution to the free-particle energy
(\ref{eq:En_FreeAssembly}) adds the term $-\zeta\sum_{\alpha=x,y,z}m_{i,\alpha}^{3}\bm{e}_{\alpha}$
to the effective field (\ref{eq:heff}) and thereby the angular frequency
of an isolated nanomagnet becomes $\omega^{\left(0\right)}=\omega_{K}\left(h+k+h_{\mathrm{d}}-\zeta\right)$.
Likewise, the corresponding dimensionless angular frequency is now
given by $\tilde{\varpi}^{\left(0\right)}=h+k+h_{\mathrm{d}}-\zeta\equiv\varpi^{\left(0\right)}+\varpi_{{\rm SE}}$
{[}see Eq. (\ref{eq:Frequency-free}){]}. We see that due to the surface
anisotropy contribution, the FMR frequency of a single nanomagnet
may either increase or decrease according to the sign of $\zeta$.
In particular, it is interesting to investigate how surface effects
may make up for the frequency red-shift induced by dipolar interactions,
as discussed earlier. Accordingly, the total frequency shift, due
to both DI and surface effects, is given by {[}see Eq. (\ref{eq:DICorrection-Disks}){]}
\begin{equation}
\Delta\varpi=-\varpi_{{\rm SE}}+\Delta\varpi_{{\rm {\rm DI}}}=\zeta-\frac{A}{\kappa^{3}}\left[\mathcal{C}_{3}+\frac{9}{16\kappa^{2}}\mathcal{C}_{5}\right].\label{eq:TotalFreqShift}
\end{equation}

For instance, for $\zeta>0$ we see that surface anisotropy may compete
with dipolar interactions. This will be discussed in Section \ref{subsec:DI+SE}.

\section{\label{sec:Results-and-discussion}Results and discussion}

Let us now discuss some of the results that can be inferred from Eq.
(\ref{eq:DICorrection-Disks}) for the effect of DI and Eq. (\ref{eq:TotalFreqShift})
when SE are included, especially in what regards the dependence of
the shift in frequency on the parameter $\kappa$, that is the ratio
of the nano-elements separation $d$ to their diameter $D$.

\subsection{Effects of dipolar interactions (ignoring surface effects)}

\begin{figure}
\begin{centering}
\includegraphics[width=0.95\columnwidth]{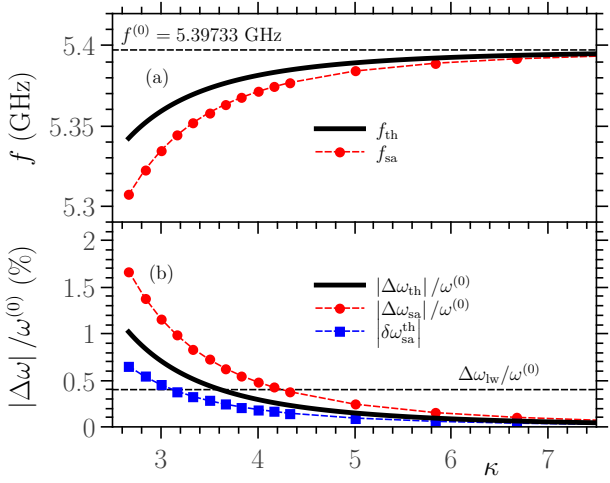} 
\par\end{centering}
\protect\caption{\foreignlanguage{american}{\label{fig:FreqShift-TheoryVSExp} (a) Resonance frequency of an interacting
$20\times20$ square array of FeV nanodisks as a function of the relative
nano-disk separation $\kappa$. The solid black line represents $f_{{\rm th}}=\left(\omega^{\left(0\right)}+\Delta\omega\right)/\left(2\pi\right)$,
where the frequency shift is obtained from Eq.~(\ref{eq:DICorrection-v2}).
In red circles we present the semi-analytical uniform mode, obtained
from the model of Ref.~\citep{fralan16jpdap}. Horizontal dashed
line represents the resonance frequency in the non-interacting case.
(b) Relative variation of the frequency shifts $\Delta\omega_{{\rm th}}/\omega^{\left(0\right)}$
and $\Delta\omega_{{\rm sa}}/\omega^{\left(0\right)}$ obtained from
theory and semi-analytical calculations, respectively. $\delta\omega_{\footnotesize{\rm sa}}^{\footnotesize{\rm th}}$
is the difference between the two approaches.}}
\end{figure}

For an order of magnitude and a comparison with other theoretical
models, we consider for instance the ferromagnetic resonance of a
finite $20\times20$ square array. We compare the results from Eq.~(\ref{eq:DICorrection-v2})
with $\mathcal{C}_{3}=7.50253$ (for a square $20\times20$ array)
and those from the semi-analytical model developed in Ref. \onlinecite{fralan16jpdap}
for multiple interacting magnetic moments. The dynamical fields arising
from the dipolar coupling, which are necessary for calculating the
FMR spectra of the nanoparticle array using the semi-analytical model,
are given in Appendix \ref{sec:ApDipolarFields}.

For the FeV disks, the thin-disk regime ($L/\left(2R\right)\ll1$)
applies and thereby the frequency shift can be calculated using Eq.~(\ref{eq:DICorrection-Disks}).
The materials parameters are \citep{mitsuzukaetal12apl,pigeauetal12prl}:
$M_{s}=1.353\times10^{6}\;\mathrm{A/m,}$ $H=1.72\;\mathrm{T},$ $K_{{\rm v}}=4.1\times10^{4}\;\mathrm{J/m^{3}}$,
and from Eq. (\ref{eq:AnisotropyField}) we can infer $H_{K}\simeq0.0606\,{\rm T}$
and $\omega_{K}=\gamma H_{K}\simeq10.67\times10^{9}\:{\rm rad}.s^{-1}$.
Next, from Fig. 3(c) of Ref. \onlinecite{pigeauetal12prl} we can
read off the frequency of the isolated elements, $f^{\left(0\right)}\simeq5.35\:{\rm GHz}$
or $\omega_{{\rm exp}}^{\left(0\right)}\simeq33.62\times10^{9}\:{\rm rad}.s^{-1}$.
We can also compute the effective field using Eq. (\ref{eq:heff}).
$\delta=R/L\simeq11.24$ leading to $N_{x}=0.956,N_{z}=0.022$ and
$H_{d}\simeq-1.59\;\mathrm{T}$. Note that for an infinitely thin
disk ($N_{x}\rightarrow1$ and $N_{z}\rightarrow0$) we would obtain
$\left|H_{d}\right|\simeq1.7\;\mathrm{T}$. Then, since $H>\left|H_{d}\right|$
we may consider the magnetic moment of the disks to be aligned along
the direction of the applied magnetic field, \emph{i.e.} $s_{x}\simeq1$
and thereby the effective field evaluates to $H_{\mathrm{eff}}\simeq0.193\;\mathrm{T}$.
This yields the (theoretical) frequency of non-interacting nano-elements
$\omega_{{\rm th}}^{\left(0\right)}=\gamma H_{\mathrm{eff}}\simeq33.91\times10^{9}\:{\rm rad}.s^{-1}$,
which is in good agreement with the experimental value $\omega_{{\rm exp}}^{\left(0\right)}$.

Now, regarding the comparison between our work and the experiments
of Ref. \onlinecite{pigeauetal12prl}, beyond the agreement of the
orders of magnitude, an important warning is necessary. In Fig. 4
of this reference, the authors plot the difference in frequency between
the anti-binding and binding modes as a function of the nano-elements
separation. Apart from the fact that only 3 values of the latter were
available, and despite the (apparent) qualitative agreement with our
theory, it is not possible to compare these experiments with our theory.
Indeed, as discussed earlier, our approach only renders the frequency
of the collective mode, which is here the binding mode, and it is
not possible to derive the frequency of the anti-binding mode as this
would require the full solution of the eigenvalue problem. On the
other hand, the individual frequencies of the two modes cannot be
extracted from these experiments because the nano-disks are not fully
identical and their distances to the sensor are not equal either.

In Fig. \ref{fig:FreqShift-TheoryVSExp}(a) we plot the resonance
frequency obtained from Eq.~(\ref{eq:DICorrection-Disks}) ($f_{{\rm th}}=\left[\omega^{\left(0\right)}+\Delta\omega\right]/\left(2\pi\right)$
- black solid line) as a function of the relative distance parameter
$\kappa$, together with the frequency ($f_{{\rm sa}}$) of the uniform
resonance mode of the system obtained from the semi-analytical model
\citep{fralan16jpdap} (red dashed line with circles). As expected,
the dipolar coupling reduces the frequency relative to the non-interacting
case (horizontal dashed line). The shift decreases as the distance
between the disks increases, which is equivalent to a decrease in
the dipolar coupling. Both theoretical calculations render the same
qualitative behavior, with some quantitative discrepancies, especially
for stronger DI. This is due to the several approximations and expansions
used in the derivation of Eq.~(\ref{eq:DICorrection-Disks}). Nonetheless,
we can clearly see that the difference is reduced as $\kappa$ increases,
reaching a good agreement for $\kappa\geq5$.

Fig. \ref{fig:FreqShift-TheoryVSExp}(b), shows the variation of the
relative frequency shift $\left|\Delta\omega\right|/\omega^{\left(0\right)}$
of each approach, and the difference $\delta\omega_{\footnotesize{\rm sa}}^{\footnotesize{\rm sh}}\equiv\left(\Delta\omega_{\footnotesize{\rm th}}-\Delta\omega_{\footnotesize{\rm sa}}\right)/\omega^{\left(0\right)}$.
$\Delta\omega_{\footnotesize{\rm th}}$ and $\Delta\omega_{\footnotesize{\rm sa}}$
are the absolute frequency shifts induced by the dipolar coupling,
obtained from Eq.~(\ref{eq:DICorrection-Disks}) and the semi-analytical
model, respectively. We see that the DI induce small frequency shifts
on the order of $1.5$ \% or even lower for the explored distances.
Furthermore, we can see that when Eq.~(\ref{eq:DICorrection-Disks})
becomes a good approximation ($\kappa\geq5$), the relative frequency
shifts are on the order of $0.3$ \%. This relative variation expressed
as a percentage is below typical relative experimental linewidths
in similar systems (linewidth $\Delta\omega_{{\rm lw}}/\left(2\pi\right)\simeq20$~MHz,
$\omega^{\left(0\right)}/\left(2\pi\right)\simeq5$~GHz \citep{pigeauetal12prl},
thus $\Delta\omega_{{\rm lw}}/\omega^{\left(0\right)}=0.4$ \%). However,
increasing $\omega^{\left(0\right)}$ (\emph{e.g.} by increasing the
applied field) reduces the relative frequency shift $\left|\Delta\omega\right|/\omega^{\left(0\right)}$
and the error $\delta\omega_{\footnotesize{\rm sa}}^{\footnotesize{\rm sh}}$.
As a consequence, the validity of our formalism {[}see Eq.~(\ref{eq:DICorrection-Disks}){]}
extends to stronger interactions (smaller $\kappa$), making it possible
to reach the regime where the predicted frequency shifts can be measured
in experiments.

\begin{figure}[h]
\begin{centering}
\includegraphics[width=0.95\columnwidth]{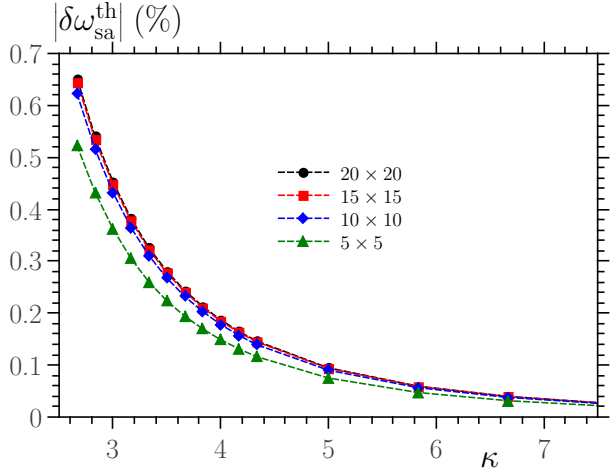} 
\par\end{centering}
\caption{\foreignlanguage{american}{\label{fig:FreqShift-ArraySize}Error in the relative shift in frequency
as a function of the relative distance $\kappa$ for different sizes
of a square array of disks. }}
\end{figure}

The dependence of $\delta\omega_{\footnotesize{\rm sa}}^{\footnotesize{\rm th}}$
on $\kappa$ for different sizes of the array is shown in Fig.~\ref{fig:FreqShift-ArraySize}.
It can be clearly seen that the error decreases for smaller arrays.
Indeed, decreasing the size of the array decreases the overall dipolar
contributions, thus making the different approximations more precise.
Furthermore, it can also be seen that the error tends to stabilize
as the size of the array increases, and no important variations are
expected for arrays larger than $20\times20$.

\subsection{\label{subsec:DI+SE}Effects of dipolar interactions including surface
effects}

\begin{figure}
\begin{centering}
\includegraphics[width=0.95\columnwidth]{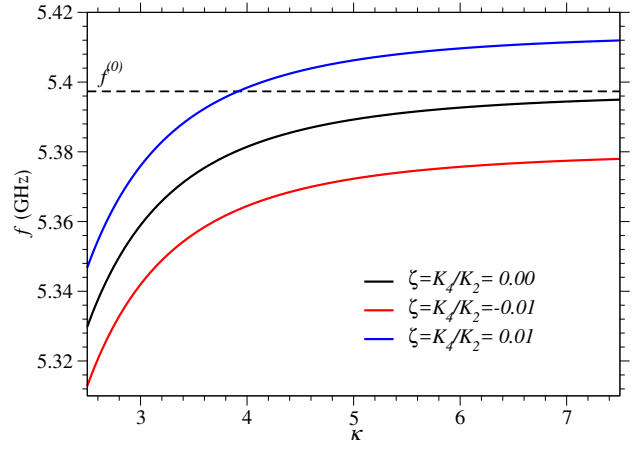} 
\par\end{centering}
\protect\caption{\foreignlanguage{american}{\label{fig:FreqShift-SE+DDI}FMR frequency as a function of the relative
nanodisk distance $\kappa$ for various surface anisotropies $\zeta=K_{4}/K_{2}$.}}
\end{figure}

Eq. (\ref{eq:TotalFreqShift}) clearly reveals a competition between
the effects of surface anisotropy and DI on the FMR frequency. In
order to better assess the role of SE, we consider an interacting
$20\times20$ array of nano-disks similar to the sample of Fig. \ref{fig:FreqShift-TheoryVSExp}.
The results are shown in Fig. \ref{fig:FreqShift-SE+DDI} where we
have restricted our investigation to the case of small surface anisotropy
\emph{i.e.} $\left|\zeta\right|\ll1$ in order to remain within the
limits of the effective macrospin model {[}see discussion in Section
\ref{sec:Surface-effects}{]}\citep{kacbon06prb,yanesetal07prb}.
First, analyzing the effect of surface anisotropy alone, we see that
changing the value and sign of $\zeta$ has a large effect on the
FMR frequency, taking $f^{\left(0\right)}$as a reference. Now, for
2D square arrays the dipolar interactions tend to maintain the magnetic
moments within the plane. In contrast, depending on the sign of $\zeta$,
SE favor a magnetic alignment along the cube facets ($\zeta<0$),
or along the cube diagonals ($\zeta>0$). Therefore, for materials
with $\zeta>0$ one may expect a competition between SE and DI. This
is what is observed in Fig. \ref{fig:FreqShift-SE+DDI}: at high densities
(small $\kappa$) DI dominate the correction to the FMR frequency
and induce a red-shift, whereas for very dilute assemblies (large
$\kappa$) each nanodisk behaves like an isolated entity and SE dominate
\citep{lavoratoetal15jpcc} and induce a blue-shift. At leading orders
in $\kappa$, the critical value $\kappa_{c}$ marking the crossover
from a red- to a blue-shift is given by $\kappa_{c}\simeq\left(A\mathcal{C}_{3}/\zeta\right)^{1/3}$.
This is the point where the blue line crosses the dashed line in Fig.
\ref{fig:FreqShift-SE+DDI}, implying that SE compensate for the DI.
For the FeV thin disks considered here, $\kappa_{c}\simeq3.9$. This
corresponds to an inter-element separation of 4 times the element
diameter, \emph{i.e. }a center-to-center distance of\emph{ $2400\,{\rm nm}$.}

The value of $\zeta$ taken here is rather small as compared to the
estimates obtained by other authors in cobalt and iron-oxide elements
\citep{urquhartetal88jap,skocoe99iop,perrai05springer}. For such
higher values (an order of magnitude larger) of surface anisotropy,
compensation of the DI effects should occur for much closer nano-elements,
or equivalently denser assemblies. However, this reasoning cannot
be taken too far, at least in the framework of our approach, since
our treatment is limited to dilute assemblies and not-too-strong surface
disorder. Nonetheless, it does confirm the screening effect of DI
by surface disorder studied earlier by the authors\cite{sabsabietal13prb,vernayetal14prb}.

\section{Conclusion}

We have developed a general formalism for deriving practical analytical
formulas for the shift in FMR frequency induced by both dipolar interactions
and surface disorder in an array of magnetic nano-elements. Even though
this has been done with the help of perturbation theory, which only
applies to relatively dilute assemblies, or equivalently for well
separated nano-elements, the general character of this formalism resides
in the fact that it applies to nano-elements of arbitrary shape and
size, and as such, it deals with the dipolar interactions beyond the
point-dipole approximation. An analytical expression for the frequency
shift induced by dipolar interactions has been explicitly derived
for an arbitrary array of monodisperse elements, and the contribution
due to their shape and size has been singled out. Next, this formalism
has been applied to the limiting case of thin disks of FeV, recently
investigated by the technique of Magnetic Resonance Force Microscopy.
We have clearly shown that the contribution of dipolar interactions
to the FMR frequency of a $2D$ array of nano-elements is a linear
function of the parameter $\xi$ which scales as the inverse of the
third power of the elements separation. In addition to this contribution,
that obtains within the point-dipole approximation, we also obtain
a contribution from the nano-elements size and shape which scales
with the inverse fifth power of the nano-elements separation. We have
also studied the effect of the array size on the frequency shift and
have found that the red-shift of the resonance is smaller for smaller
arrays. The effects of surface anisotropy on the frequency shift have
been taken into account with the help of an effective macroscopic
model for the isolated nano-elements. Depending on the sign of the
corresponding contribution, which changes with the properties pertaining
to the nano-element itself, we may obtain either a blue-shift or a
red-shift of the FMR frequency. Correspondingly, this may lead to
a competition or a concomitant effect with the dipolar interactions.
This means that surface anisotropy and dipolar interactions provide
us with a handle for adjusting the resonance frequency of nano-magnet
assemblies. 
\begin{acknowledgments}
A. F. Franco acknowledges financial support from the FONDECYT postdoctoral
project No 3150180. 
\end{acknowledgments}

\appendix

\section{\label{sec:The-pseudo-Hessian-matrix}The pseudo-Hessian matrix elements
for the DI contribution}

The following general expressions \citep{BastardisEtal-jpcm2017}
are used in the calculation of the second derivatives of the DI contribution
with respect to the angular variables $\left(\theta_{i},\varphi_{i}\right)$

\begin{eqnarray}
\partial_{\theta_{i}\theta_{k}}^{2}\mathcal{E}_{{\rm DI}} & = & -\delta_{ik}\sum_{j=1}^{\mathcal{N}}\xi_{ij}\left(1-\delta_{ij}\right)\mathbf{s}_{i}\cdot\mathcal{D}_{ij}\cdot\mathbf{s}_{j}\label{eq:SecondDerivsDI}\\
 &  & +\xi_{ik}\left(1-\delta_{ik}\right)\left(\mathbf{e}_{\theta_{i}}\cdot\mathcal{D}_{ik}\cdot\mathbf{e}_{\theta_{k}}\right)\nonumber \\
\partial_{\varphi_{i}\varphi_{k}}^{2}\mathcal{E}_{{\rm DI}} & = & -\delta_{ik}\sum_{j=1}^{\mathcal{N}}\xi_{ij}\left(1-\delta_{ij}\right)\times\nonumber \\
 &  & \left[\left(\sin^{2}\theta_{i}\,\mathbf{s}_{i}+\sin\theta_{i}\cos\theta_{i}\,\mathbf{e}_{\theta_{i}}\right)\cdot\mathcal{D}_{ij}\cdot\mathbf{s}_{j}\right]\nonumber \\
 &  & +\xi_{ik}\left(1-\delta_{ik}\right)\sin\theta_{i}\sin\theta_{k}\left(\mathbf{e}_{\varphi_{i}}\cdot\mathcal{D}_{ik}\cdot\mathbf{e}_{\varphi_{k}}\right)\nonumber \\
\partial_{\theta_{k}\varphi_{i}}^{2}\mathcal{E}_{{\rm DI}} & = & \delta_{ik}\sum_{j=1}^{\mathcal{N}}\xi_{ij}\left(1-\delta_{ij}\right)\cos\theta_{i}\left(\mathbf{e}_{\varphi_{i}}\cdot\mathcal{D}_{ij}\cdot\mathbf{s}_{j}\right)\nonumber \\
 &  & +\xi_{ik}\left(1-\delta_{ik}\right)\sin\theta_{i}\left(\mathbf{e}_{\varphi_{i}}\cdot\mathcal{D}_{ik}\cdot\mathbf{e}_{\theta_{k}}\right)\nonumber \\
\partial_{\varphi_{k}\theta_{i}}^{2}\mathcal{E}_{{\rm DI}} & = & \delta_{ik}\sum_{j=1}^{\mathcal{N}}\xi_{ij}\left(1-\delta_{ij}\right)\cos\theta_{i}\left(\mathbf{e}_{\varphi_{i}}\cdot\mathcal{D}_{ij}\cdot\mathbf{s}_{j}\right)\nonumber \\
 &  & +\xi_{ik}\left(1-\delta_{ik}\right)\sin\theta_{k}\left(\mathbf{e}_{\theta_{i}}\cdot\mathcal{D}_{ik}\cdot\mathbf{e}_{\varphi_{k}}\right),\nonumber 
\end{eqnarray}
with

\begin{eqnarray*}
\mathbf{e}_{\theta_{i}} & = & \partial_{\theta_{i}}\mathbf{s}_{i}=\left(\begin{array}{c}
\cos\theta_{i}\cos\varphi_{i}\\
\cos\theta_{i}\sin\varphi_{i}\\
-\sin\theta_{i}
\end{array}\right),\\
\mathbf{e}_{\varphi_{i}} & = & \frac{1}{\sin\theta_{i}}\partial_{\varphi_{i}}\mathbf{s}_{i}=\left(\begin{array}{c}
-\sin\varphi_{i}\\
\cos\varphi_{i}\\
0
\end{array}\right).
\end{eqnarray*}

\section{\label{sec:ApDipolarFields}Dynamical dipolar fields}

The dynamical fields due to the dipolar coupling can be calculated
in the context of the model presented in \citep{fralan16jpdap} for
the propagation of in-plane spin waves in multilayer systems. This
model can be applied to the array of interacting nano-particles presented
here for the zero wave-vector (uniform mode). Starting from Eq.~(\ref{eq:HorizDimerEnergy})
(in SI units), and following a similar procedure as the one presented
in \citep{fralan16jpdap}, the following dipolar dynamical fields
were obtained

\begin{eqnarray*}
H_{x_{i}x_{i}}^{d} & = & -\sum_{j\neq i}\frac{M_{s}^{j}}{4\pi r_{ij}^{3}}\left[\Phi_{ij}\cos\theta_{i}\cos\theta_{j}+\sin\theta_{i}\sin\theta_{j}\right.\\
 &  & \times\left(\cos\varphi_{i,j}^{-}-\left(2+\Phi_{ij}\right)\left(r_{ij,x}r_{ij,y}\sin\varphi_{i,j}^{+}\right.\right.\\
 &  & \left.\left.\left.+r_{ij,x}^{2}\cos\varphi_{i}\cos\varphi_{j}+r_{ij,j}^{2}\sin\varphi_{i}\sin\varphi_{j}\right)\right)\right]\\
H_{y_{i}y_{i}}^{d} & = & H_{x_{i}x_{i}}^{d}\\
H_{x_{i}x_{j}}^{d} & = & \frac{M_{s}^{j}}{4\pi r_{ij}^{3}}\left[\Phi_{ij}\sin\theta_{i}\sin\theta_{j}+\cos\theta_{i}\cos\theta_{j}\right.\\
 &  & \times\left(\cos\varphi_{i,j}^{-}-\left(2+\Phi_{ij}\right)\left(r_{ij,x}r_{ij,y}\sin\varphi_{i,j}^{+}\right.\right.\\
 &  & \left.\left.\left.+r_{ij,x}^{2}\cos\varphi_{i}\cos\varphi_{j}+r_{ij,y}^{2}\sin\varphi_{i}\sin\varphi_{j}\right)\right)\right]\\
H_{y_{i}y_{j}}^{d} & = & \frac{M_{s}^{j}}{4\pi r_{ij}^{3}}\left[\cos\varphi_{i,j}^{-}-\left(2+\Phi_{ij}\right)\left(r_{ij,x}^{2}\sin\varphi_{i}\sin\varphi_{j}\right.\right.\\
 &  & \left.\left.+r_{ij,y}^{2}\cos\varphi_{i}\cos\varphi_{j}-r_{ij,x}r_{ij,y}\sin\varphi_{i,j}^{-}\right)\right]\\
H_{x_{i}y_{j}}^{d} & = & \frac{M_{s}^{j}}{4\pi r_{ij}^{3}}\cos\theta_{i}\left[\sin\varphi_{i,j}^{-}+\left(2+\Phi_{ij}\right)\right.\\
 &  & \times\left(r_{ij,x}^{2}\cos\varphi_{i}\sin\varphi_{j}-r_{ij,y}^{2}\sin\varphi_{i}\cos\varphi_{j}\right.\\
 &  & \left.\left.-r_{ij,x}r_{ij,y}\cos\varphi_{i,j}^{+}\right)\right]\\
H_{y_{i}x_{j}}^{d} & = & -\frac{M_{s}^{j}}{4\pi r_{ij}^{3}}\cos\theta_{j}\left[\sin\varphi_{i,j}^{-}-\left(2+\Phi_{ij}\right)\right.\\
 &  & \times\left(r_{ij,x}^{2}\sin\varphi_{i}\cos\varphi_{j}-r_{ij,y}^{2}\cos\varphi_{i}\sin\varphi_{j}\right.\\
 &  & \left.\left.-r_{ij,x}r_{ij,y}\cos\varphi_{i,j}^{+}\right)\right],
\end{eqnarray*}
with $\varphi_{i,j}^{\pm}\equiv\varphi_{i}\pm\varphi_{j}$.

\bibliography{hkbib}

\begin{thebibliography}{44}
\expandafter\ifx\csname natexlab\endcsname\relax\def\natexlab#1{#1}\fi
\expandafter\ifx\csname bibnamefont\endcsname\relax
  \def\bibnamefont#1{#1}\fi
\expandafter\ifx\csname bibfnamefont\endcsname\relax
  \def\bibfnamefont#1{#1}\fi
\expandafter\ifx\csname citenamefont\endcsname\relax
  \def\citenamefont#1{#1}\fi
\expandafter\ifx\csname url\endcsname\relax
  \def\url#1{\texttt{#1}}\fi
\expandafter\ifx\csname urlprefix\endcsname\relax\def\urlprefix{URL }\fi
\providecommand{\bibinfo}[2]{#2}
\providecommand{\eprint}[2][]{\url{#2}}

\bibitem[{\citenamefont{Vonsovskii}(1966)}]{vonsovskii66pp}
\bibinfo{author}{\bibfnamefont{S.~V.} \bibnamefont{Vonsovskii}},
  \emph{\bibinfo{title}{{Ferromagnetic Resonance: The Phenomenon of Resonant
  Absorption of a High-Frequency Magnetic Field in Ferromagnetic Substances}}}
  (\bibinfo{publisher}{Pergamon Press, Oxford}, \bibinfo{year}{1966}).

\bibitem[{\citenamefont{{A.G. Gurevich and G.A.
  Melkov}}(1996)}]{gurmel96crcpress}
\bibinfo{author}{\bibnamefont{{A.G. Gurevich and G.A. Melkov}}},
  \emph{\bibinfo{title}{{Magnetization oscillations and waves}}}
  (\bibinfo{publisher}{CSC Press}, \bibinfo{address}{Florida},
  \bibinfo{year}{1996}).

\bibitem[{\citenamefont{Tran}(2006)}]{tran06phd}
\bibinfo{author}{\bibfnamefont{M.}~\bibnamefont{Tran}}, Master's thesis,
  \bibinfo{school}{Institut National des Sciences Appliquees de Toulouse},
  \bibinfo{address}{Toulouse} (\bibinfo{year}{2006}).

\bibitem[{\citenamefont{{B. Heinrich}}(1994)}]{heinrich94springer}
\bibinfo{author}{\bibnamefont{{B. Heinrich}}}, in
  \emph{\bibinfo{booktitle}{{Ultrathin magnetic structures II}}}, edited by
  \bibinfo{editor}{\bibfnamefont{B.}~\bibnamefont{Heinrich}} \bibnamefont{and}
  \bibinfo{editor}{\bibfnamefont{J.}~\bibnamefont{Bland}}
  (\bibinfo{publisher}{Springer-Verlag}, \bibinfo{address}{Berlin},
  \bibinfo{year}{1994}), p. \bibinfo{pages}{195}.

\bibitem[{\citenamefont{{J. S. Lee, R. P. Tan, J. H. Wu, and Y. K.
  Kim}}(2011)}]{leeetal11apl}
\bibinfo{author}{\bibnamefont{{J. S. Lee, R. P. Tan, J. H. Wu, and Y. K.
  Kim}}}, \bibinfo{journal}{Appl. Phys. Lett.} \textbf{\bibinfo{volume}{99}},
  \bibinfo{pages}{062506} (\bibinfo{year}{2011}),
  \eprint{https://doi.org/10.1063/1.3624833},
  \urlprefix\url{https://doi.org/10.1063/1.3624833}.

\bibitem[{\citenamefont{Gon{\c{c}}alves
  et~al.}(2013)\citenamefont{Gon{\c{c}}alves, Barsukov, Chen, Yang, Katine, and
  Krivorotov}}]{goncalvesetal13apl}
\bibinfo{author}{\bibfnamefont{A.~M.} \bibnamefont{Gon{\c{c}}alves}},
  \bibinfo{author}{\bibfnamefont{I.}~\bibnamefont{Barsukov}},
  \bibinfo{author}{\bibfnamefont{Y.-J.} \bibnamefont{Chen}},
  \bibinfo{author}{\bibfnamefont{L.}~\bibnamefont{Yang}},
  \bibinfo{author}{\bibfnamefont{J.~A.} \bibnamefont{Katine}},
  \bibnamefont{and} \bibinfo{author}{\bibfnamefont{I.~N.}
  \bibnamefont{Krivorotov}}, \bibinfo{journal}{Applied Physics Letters}
  \textbf{\bibinfo{volume}{103}}, \bibinfo{pages}{172406}
  (\bibinfo{year}{2013}), \eprint{https://doi.org/10.1063/1.4826927},
  \urlprefix\url{https://doi.org/10.1063/1.4826927}.

\bibitem[{\citenamefont{Schoeppner et~al.}(2014)\citenamefont{Schoeppner,
  Wagner, Stienen, Meckenstock, Farle, Narkowicz, Suter, and
  Lindner}}]{Schoeppneretal14jap}
\bibinfo{author}{\bibfnamefont{C.}~\bibnamefont{Schoeppner}},
  \bibinfo{author}{\bibfnamefont{K.}~\bibnamefont{Wagner}},
  \bibinfo{author}{\bibfnamefont{S.}~\bibnamefont{Stienen}},
  \bibinfo{author}{\bibfnamefont{R.}~\bibnamefont{Meckenstock}},
  \bibinfo{author}{\bibfnamefont{M.}~\bibnamefont{Farle}},
  \bibinfo{author}{\bibfnamefont{R.}~\bibnamefont{Narkowicz}},
  \bibinfo{author}{\bibfnamefont{D.}~\bibnamefont{Suter}}, \bibnamefont{and}
  \bibinfo{author}{\bibfnamefont{J.}~\bibnamefont{Lindner}},
  \bibinfo{journal}{Journal of Applied Physics} \textbf{\bibinfo{volume}{116}},
  \bibinfo{pages}{033913} (\bibinfo{year}{2014}),
  \eprint{https://doi.org/10.1063/1.4890515},
  \urlprefix\url{https://doi.org/10.1063/1.4890515}.

\bibitem[{\citenamefont{Ollefs et~al.}(2015)\citenamefont{Ollefs, Meckenstock,
  Spoddig, R{\"o}mer, Hassel, Sch{\"o}ppner, Ney, Farle, and
  Ney}}]{ollefsetal15jap}
\bibinfo{author}{\bibfnamefont{K.}~\bibnamefont{Ollefs}},
  \bibinfo{author}{\bibfnamefont{R.}~\bibnamefont{Meckenstock}},
  \bibinfo{author}{\bibfnamefont{D.}~\bibnamefont{Spoddig}},
  \bibinfo{author}{\bibfnamefont{F.~M.} \bibnamefont{R{\"o}mer}},
  \bibinfo{author}{\bibfnamefont{C.}~\bibnamefont{Hassel}},
  \bibinfo{author}{\bibfnamefont{C.}~\bibnamefont{Sch{\"o}ppner}},
  \bibinfo{author}{\bibfnamefont{V.}~\bibnamefont{Ney}},
  \bibinfo{author}{\bibfnamefont{M.}~\bibnamefont{Farle}}, \bibnamefont{and}
  \bibinfo{author}{\bibfnamefont{A.}~\bibnamefont{Ney}},
  \bibinfo{journal}{Journal of Applied Physics} \textbf{\bibinfo{volume}{117}},
  \bibinfo{pages}{223906} (\bibinfo{year}{2015}),
  \eprint{https://doi.org/10.1063/1.4922248},
  \urlprefix\url{https://doi.org/10.1063/1.4922248}.

\bibitem[{\citenamefont{Sidles et~al.}(1995)\citenamefont{Sidles, Garbini,
  Bruland, Rugar, Z{\"u}ger, Hoen, and Yannoni}}]{Sidles_RevModPhys}
\bibinfo{author}{\bibfnamefont{J.~A.} \bibnamefont{Sidles}},
  \bibinfo{author}{\bibfnamefont{J.~L.} \bibnamefont{Garbini}},
  \bibinfo{author}{\bibfnamefont{K.~J.} \bibnamefont{Bruland}},
  \bibinfo{author}{\bibfnamefont{D.}~\bibnamefont{Rugar}},
  \bibinfo{author}{\bibfnamefont{O.}~\bibnamefont{Z{\"u}ger}},
  \bibinfo{author}{\bibfnamefont{S.}~\bibnamefont{Hoen}}, \bibnamefont{and}
  \bibinfo{author}{\bibfnamefont{C.~S.} \bibnamefont{Yannoni}},
  \bibinfo{journal}{Rev. Mod. Phys.} \textbf{\bibinfo{volume}{67}},
  \bibinfo{pages}{249} (\bibinfo{year}{1995}),
  \urlprefix\url{http://link.aps.org/doi/10.1103/RevModPhys.67.249}.

\bibitem[{\citenamefont{{Lavenant, H., Naletov, V.~V. Klein, O. De Loubens, G.
  Laura, C. De Teresa, J.~M.}}(2014)}]{lavenantetal2014nanofab}
\bibinfo{author}{\bibnamefont{{Lavenant, H., Naletov, V.~V. Klein, O. De
  Loubens, G. Laura, C. De Teresa, J.~M.}}}, \bibinfo{journal}{Nanofabrication}
  \textbf{\bibinfo{volume}{1}}, \bibinfo{pages}{65} (\bibinfo{year}{2014}).

\bibitem[{\citenamefont{Kittel}(1948)}]{kittel48pr}
\bibinfo{author}{\bibfnamefont{C.}~\bibnamefont{Kittel}},
  \bibinfo{journal}{Phys. Rev.} \textbf{\bibinfo{volume}{73}},
  \bibinfo{pages}{155} (\bibinfo{year}{1948}),
  \urlprefix\url{https://link.aps.org/doi/10.1103/PhysRev.73.155}.

\bibitem[{\citenamefont{Strijkers et~al.}(1999)\citenamefont{Strijkers,
  Dalderop, Broeksteeg, Swagten, and de~Jonge}}]{strijkersetal99jap}
\bibinfo{author}{\bibfnamefont{G.~J.} \bibnamefont{Strijkers}},
  \bibinfo{author}{\bibfnamefont{J.~H.~J.} \bibnamefont{Dalderop}},
  \bibinfo{author}{\bibfnamefont{M.~A.~A.} \bibnamefont{Broeksteeg}},
  \bibinfo{author}{\bibfnamefont{H.~J.~M.} \bibnamefont{Swagten}},
  \bibnamefont{and} \bibinfo{author}{\bibfnamefont{W.~J.~M.}
  \bibnamefont{de~Jonge}}, \bibinfo{journal}{Journal of Applied Physics}
  \textbf{\bibinfo{volume}{86}}, \bibinfo{pages}{5141} (\bibinfo{year}{1999}),
  \urlprefix\url{http://scitation.aip.org/content/aip/journal/jap/86/9/10.1063/1.371490}.

\bibitem[{\citenamefont{Encinas-Oropesa
  et~al.}(2001)\citenamefont{Encinas-Oropesa, Demand, Piraux, Ebels, and
  Huynen}}]{ancinasetal01jap}
\bibinfo{author}{\bibfnamefont{A.}~\bibnamefont{Encinas-Oropesa}},
  \bibinfo{author}{\bibfnamefont{M.}~\bibnamefont{Demand}},
  \bibinfo{author}{\bibfnamefont{L.}~\bibnamefont{Piraux}},
  \bibinfo{author}{\bibfnamefont{U.}~\bibnamefont{Ebels}}, \bibnamefont{and}
  \bibinfo{author}{\bibfnamefont{I.}~\bibnamefont{Huynen}},
  \bibinfo{journal}{Journal of Applied Physics} \textbf{\bibinfo{volume}{89}},
  \bibinfo{pages}{6704} (\bibinfo{year}{2001}),
  \eprint{https://doi.org/10.1063/1.1362638},
  \urlprefix\url{https://doi.org/10.1063/1.1362638}.

\bibitem[{\citenamefont{Yang et~al.}(2009)\citenamefont{Yang, Hasegawa,
  Takahashi, and Ogawa}}]{Yang_etal_APL2009}
\bibinfo{author}{\bibfnamefont{H.~T.} \bibnamefont{Yang}},
  \bibinfo{author}{\bibfnamefont{D.}~\bibnamefont{Hasegawa}},
  \bibinfo{author}{\bibfnamefont{M.}~\bibnamefont{Takahashi}},
  \bibnamefont{and} \bibinfo{author}{\bibfnamefont{T.}~\bibnamefont{Ogawa}},
  \bibinfo{journal}{Applied Physics Letters} \textbf{\bibinfo{volume}{94}},
  \bibinfo{pages}{013103} (\bibinfo{year}{2009}),
  \eprint{https://doi.org/10.1063/1.3063032},
  \urlprefix\url{https://doi.org/10.1063/1.3063032}.

\bibitem[{\citenamefont{Urquhart et~al.}(1988)\citenamefont{Urquhart, Heinrich,
  Cochran, Arrott, and Myrtle}}]{urquhartetal88jap}
\bibinfo{author}{\bibfnamefont{K.~B.} \bibnamefont{Urquhart}},
  \bibinfo{author}{\bibfnamefont{B.}~\bibnamefont{Heinrich}},
  \bibinfo{author}{\bibfnamefont{J.~F.} \bibnamefont{Cochran}},
  \bibinfo{author}{\bibfnamefont{A.~S.} \bibnamefont{Arrott}},
  \bibnamefont{and} \bibinfo{author}{\bibfnamefont{K.}~\bibnamefont{Myrtle}},
  \bibinfo{journal}{Journal of Applied Physics} \textbf{\bibinfo{volume}{64}},
  \bibinfo{pages}{5334} (\bibinfo{year}{1988}),
  \eprint{https://doi.org/10.1063/1.342362},
  \urlprefix\url{https://doi.org/10.1063/1.342362}.

\bibitem[{\citenamefont{{R. Skomski and J.M.D. Coey}}(1999)}]{skocoe99iop}
\bibinfo{author}{\bibnamefont{{R. Skomski and J.M.D. Coey}}},
  \emph{\bibinfo{title}{{Permanent Magnetism, Studies in Condensed Matter
  Physics Vol. 1}}} (\bibinfo{publisher}{IOP Publishing},
  \bibinfo{address}{London}, \bibinfo{year}{1999}).

\bibitem[{\citenamefont{{R. Perzynski and Yu.L.
  Raikher}}(2005)}]{perrai05springer}
\bibinfo{author}{\bibnamefont{{R. Perzynski and Yu.L. Raikher}}}, in
  \emph{\bibinfo{booktitle}{{Surface effects in magnetic nanoparticles}}},
  edited by \bibinfo{editor}{\bibfnamefont{D.}~\bibnamefont{Fiorani}}
  (\bibinfo{publisher}{Springer}, \bibinfo{address}{Berlin},
  \bibinfo{year}{2005}), p. \bibinfo{pages}{141}.

\bibitem[{\citenamefont{Kachkachi and Bonet}(2006)}]{kacbon06prb}
\bibinfo{author}{\bibfnamefont{H.}~\bibnamefont{Kachkachi}} \bibnamefont{and}
  \bibinfo{author}{\bibfnamefont{E.}~\bibnamefont{Bonet}},
  \bibinfo{journal}{Phys. Rev. B} \textbf{\bibinfo{volume}{73}},
  \bibinfo{pages}{224402} (\bibinfo{year}{2006}),
  \urlprefix\url{https://link.aps.org/doi/10.1103/PhysRevB.73.224402}.

\bibitem[{\citenamefont{Yanes et~al.}(2007)\citenamefont{Yanes,
  Chubykalo-Fesenko, Kachkachi, Garanin, Evans, and
  Chantrell}}]{yanesetal07prb}
\bibinfo{author}{\bibfnamefont{R.}~\bibnamefont{Yanes}},
  \bibinfo{author}{\bibfnamefont{O.}~\bibnamefont{Chubykalo-Fesenko}},
  \bibinfo{author}{\bibfnamefont{H.}~\bibnamefont{Kachkachi}},
  \bibinfo{author}{\bibfnamefont{D.~A.} \bibnamefont{Garanin}},
  \bibinfo{author}{\bibfnamefont{R.}~\bibnamefont{Evans}}, \bibnamefont{and}
  \bibinfo{author}{\bibfnamefont{R.~W.} \bibnamefont{Chantrell}},
  \bibinfo{journal}{Phys. Rev. B} \textbf{\bibinfo{volume}{76}},
  \bibinfo{pages}{064416} (\bibinfo{year}{2007}),
  \urlprefix\url{https://link.aps.org/doi/10.1103/PhysRevB.76.064416}.

\bibitem[{\citenamefont{Garanin and Kachkachi}(2003)}]{garkac03prl}
\bibinfo{author}{\bibfnamefont{D.~A.} \bibnamefont{Garanin}} \bibnamefont{and}
  \bibinfo{author}{\bibfnamefont{H.}~\bibnamefont{Kachkachi}},
  \bibinfo{journal}{Phys. Rev. Lett.} \textbf{\bibinfo{volume}{90}},
  \bibinfo{pages}{065504} (\bibinfo{year}{2003}),
  \urlprefix\url{http://link.aps.org/doi/10.1103/PhysRevLett.90.065504}.

\bibitem[{\citenamefont{Kachkachi}(2007)}]{kachkachi07j3m}
\bibinfo{author}{\bibfnamefont{H.}~\bibnamefont{Kachkachi}},
  \bibinfo{journal}{{Journal of Magnetism and Magnetic Materials}}
  \textbf{\bibinfo{volume}{316}}, \bibinfo{pages}{248 } (\bibinfo{year}{2007}),
  ISSN \bibinfo{issn}{0304-8853}, \bibinfo{note}{{Proceedings of the Joint
  European Magnetic Symposia}},
  \urlprefix\url{http://www.sciencedirect.com/science/article/pii/S0304885307005252}.

\bibitem[{\citenamefont{Verba et~al.}(2012)\citenamefont{Verba, Melkov,
  Tiberkevich, and Slavin}}]{verbaetal12prb}
\bibinfo{author}{\bibfnamefont{R.}~\bibnamefont{Verba}},
  \bibinfo{author}{\bibfnamefont{G.}~\bibnamefont{Melkov}},
  \bibinfo{author}{\bibfnamefont{V.}~\bibnamefont{Tiberkevich}},
  \bibnamefont{and} \bibinfo{author}{\bibfnamefont{A.}~\bibnamefont{Slavin}},
  \bibinfo{journal}{Phys. Rev. B} \textbf{\bibinfo{volume}{85}},
  \bibinfo{pages}{014427} (\bibinfo{year}{2012}),
  \urlprefix\url{https://link.aps.org/doi/10.1103/PhysRevB.85.014427}.

\bibitem[{\citenamefont{Beleggia et~al.}(2004)\citenamefont{Beleggia, Tandon,
  Zhu, and Graef}}]{beleggiaetal04jmmm278}
\bibinfo{author}{\bibfnamefont{M.}~\bibnamefont{Beleggia}},
  \bibinfo{author}{\bibfnamefont{S.}~\bibnamefont{Tandon}},
  \bibinfo{author}{\bibfnamefont{Y.}~\bibnamefont{Zhu}}, \bibnamefont{and}
  \bibinfo{author}{\bibfnamefont{M.~D.} \bibnamefont{Graef}},
  \bibinfo{journal}{Journal of Magnetism and Magnetic Materials}
  \textbf{\bibinfo{volume}{278}}, \bibinfo{pages}{270 } (\bibinfo{year}{2004}),
  ISSN \bibinfo{issn}{0304-8853},
  \urlprefix\url{http://www.sciencedirect.com/science/article/pii/S0304885304000186}.

\bibitem[{\citenamefont{Naletov et~al.}(2011)\citenamefont{Naletov, de~Loubens,
  Albuquerque, Borlenghi, Cros, Faini, Grollier, Hurdequint, Locatelli, Pigeau
  et~al.}}]{naletovetal11prb}
\bibinfo{author}{\bibfnamefont{V.~V.} \bibnamefont{Naletov}},
  \bibinfo{author}{\bibfnamefont{G.}~\bibnamefont{de~Loubens}},
  \bibinfo{author}{\bibfnamefont{G.}~\bibnamefont{Albuquerque}},
  \bibinfo{author}{\bibfnamefont{S.}~\bibnamefont{Borlenghi}},
  \bibinfo{author}{\bibfnamefont{V.}~\bibnamefont{Cros}},
  \bibinfo{author}{\bibfnamefont{G.}~\bibnamefont{Faini}},
  \bibinfo{author}{\bibfnamefont{J.}~\bibnamefont{Grollier}},
  \bibinfo{author}{\bibfnamefont{H.}~\bibnamefont{Hurdequint}},
  \bibinfo{author}{\bibfnamefont{N.}~\bibnamefont{Locatelli}},
  \bibinfo{author}{\bibfnamefont{B.}~\bibnamefont{Pigeau}},
  \bibnamefont{et~al.}, \bibinfo{journal}{Phys. Rev. B}
  \textbf{\bibinfo{volume}{84}}, \bibinfo{pages}{224423}
  (\bibinfo{year}{2011}),
  \urlprefix\url{https://link.aps.org/doi/10.1103/PhysRevB.84.224423}.

\bibitem[{\citenamefont{Sukhov et~al.}(2014)\citenamefont{Sukhov, Horley,
  Berakdar, Terwey, Meckenstock, and Farle}}]{sukhovetal14ieee}
\bibinfo{author}{\bibfnamefont{A.}~\bibnamefont{Sukhov}},
  \bibinfo{author}{\bibfnamefont{P.~P.} \bibnamefont{Horley}},
  \bibinfo{author}{\bibfnamefont{J.}~\bibnamefont{Berakdar}},
  \bibinfo{author}{\bibfnamefont{A.}~\bibnamefont{Terwey}},
  \bibinfo{author}{\bibfnamefont{R.}~\bibnamefont{Meckenstock}},
  \bibnamefont{and} \bibinfo{author}{\bibfnamefont{M.}~\bibnamefont{Farle}},
  \bibinfo{journal}{IEEE Transactions on Magnetics}
  \textbf{\bibinfo{volume}{50}}, \bibinfo{pages}{1} (\bibinfo{year}{2014}),
  ISSN \bibinfo{issn}{0018-9464},
  \urlprefix\url{http://ieeexplore.ieee.org/stamp/stamp.jsp?tp=&arnumber=6828754&isnumber=6980152}.

\bibitem[{\citenamefont{Shendruk et~al.}(2007)\citenamefont{Shendruk,
  Desautels, Southern, and van Lierop}}]{shendruketal07nanotech}
\bibinfo{author}{\bibfnamefont{T.~N.} \bibnamefont{Shendruk}},
  \bibinfo{author}{\bibfnamefont{R.~D.} \bibnamefont{Desautels}},
  \bibinfo{author}{\bibfnamefont{B.~W.} \bibnamefont{Southern}},
  \bibnamefont{and} \bibinfo{author}{\bibfnamefont{J.}~\bibnamefont{van
  Lierop}}, \bibinfo{journal}{Nanotechnology} \textbf{\bibinfo{volume}{18}},
  \bibinfo{pages}{455704} (\bibinfo{year}{2007}),
  \urlprefix\url{http://stacks.iop.org/0957-4484/18/i=45/a=455704}.

\bibitem[{\citenamefont{Mitsuzuka et~al.}(2012)\citenamefont{Mitsuzuka, Lacour,
  Hehn, Andrieu, and Montaigne}}]{mitsuzukaetal12apl}
\bibinfo{author}{\bibfnamefont{K.}~\bibnamefont{Mitsuzuka}},
  \bibinfo{author}{\bibfnamefont{D.}~\bibnamefont{Lacour}},
  \bibinfo{author}{\bibfnamefont{M.}~\bibnamefont{Hehn}},
  \bibinfo{author}{\bibfnamefont{S.}~\bibnamefont{Andrieu}}, \bibnamefont{and}
  \bibinfo{author}{\bibfnamefont{F.}~\bibnamefont{Montaigne}},
  \bibinfo{journal}{Applied Physics Letters} \textbf{\bibinfo{volume}{100}},
  \bibinfo{pages}{192406} (\bibinfo{year}{2012}),
  \eprint{https://doi.org/10.1063/1.4711219},
  \urlprefix\url{https://doi.org/10.1063/1.4711219}.

\bibitem[{\citenamefont{Lisiecki and Nakamae}(2014)}]{lisnak13icfpm8}
\bibinfo{author}{\bibfnamefont{I.}~\bibnamefont{Lisiecki}} \bibnamefont{and}
  \bibinfo{author}{\bibfnamefont{S.}~\bibnamefont{Nakamae}},
  \bibinfo{journal}{Journal of Physics: Conference Series}
  \textbf{\bibinfo{volume}{521}}, \bibinfo{pages}{012007}
  (\bibinfo{year}{2014}),
  \urlprefix\url{http://stacks.iop.org/1742-6596/521/i=1/a=012007}.

\bibitem[{\citenamefont{Khusrhid et~al.}(2014)\citenamefont{Khusrhid,
  Porshokouh, Phan, Mukherjee, and Srikanth}}]{khusrhidetal14jap}
\bibinfo{author}{\bibfnamefont{H.}~\bibnamefont{Khusrhid}},
  \bibinfo{author}{\bibfnamefont{Z.~N.} \bibnamefont{Porshokouh}},
  \bibinfo{author}{\bibfnamefont{M.-H.} \bibnamefont{Phan}},
  \bibinfo{author}{\bibfnamefont{P.}~\bibnamefont{Mukherjee}},
  \bibnamefont{and} \bibinfo{author}{\bibfnamefont{H.}~\bibnamefont{Srikanth}},
  \bibinfo{journal}{Journal of Applied Physics} \textbf{\bibinfo{volume}{115}},
  \bibinfo{pages}{17E131} (\bibinfo{year}{2014}),
  \eprint{https://doi.org/10.1063/1.4868619},
  \urlprefix\url{https://doi.org/10.1063/1.4868619}.

\bibitem[{\citenamefont{Lavorato et~al.}(2015)\citenamefont{Lavorato, Peddis,
  Lima, Troiani, Agostinelli, Fiorani, Zysler, and
  Winkler}}]{lavoratoetal15jpcc}
\bibinfo{author}{\bibfnamefont{G.~C.} \bibnamefont{Lavorato}},
  \bibinfo{author}{\bibfnamefont{D.}~\bibnamefont{Peddis}},
  \bibinfo{author}{\bibfnamefont{E.}~\bibnamefont{Lima}},
  \bibinfo{author}{\bibfnamefont{H.~E.} \bibnamefont{Troiani}},
  \bibinfo{author}{\bibfnamefont{E.}~\bibnamefont{Agostinelli}},
  \bibinfo{author}{\bibfnamefont{D.}~\bibnamefont{Fiorani}},
  \bibinfo{author}{\bibfnamefont{R.~D.} \bibnamefont{Zysler}},
  \bibnamefont{and} \bibinfo{author}{\bibfnamefont{E.~L.}
  \bibnamefont{Winkler}}, \bibinfo{journal}{The Journal of Physical Chemistry
  C} \textbf{\bibinfo{volume}{119}}, \bibinfo{pages}{15755}
  (\bibinfo{year}{2015}), \eprint{https://doi.org/10.1021/acs.jpcc.5b04448},
  \urlprefix\url{https://doi.org/10.1021/acs.jpcc.5b04448}.

\bibitem[{\citenamefont{Tandon et~al.}(2004)\citenamefont{Tandon, Beleggia,
  Zhu, and Graef}}]{tandonetalI04jmmm}
\bibinfo{author}{\bibfnamefont{S.}~\bibnamefont{Tandon}},
  \bibinfo{author}{\bibfnamefont{M.}~\bibnamefont{Beleggia}},
  \bibinfo{author}{\bibfnamefont{Y.}~\bibnamefont{Zhu}}, \bibnamefont{and}
  \bibinfo{author}{\bibfnamefont{M.~D.} \bibnamefont{Graef}},
  \bibinfo{journal}{Journal of Magnetism and Magnetic Materials}
  \textbf{\bibinfo{volume}{271}}, \bibinfo{pages}{9 } (\bibinfo{year}{2004}),
  ISSN \bibinfo{issn}{0304-8853},
  \urlprefix\url{http://www.sciencedirect.com/science/article/pii/S0304885303007467}.

\bibitem[{\citenamefont{Caciagli et~al.}(2018)\citenamefont{Caciagli, Baars,
  Philipse, and Kuipers}}]{Caciagli_JMMM_2018}
\bibinfo{author}{\bibfnamefont{A.}~\bibnamefont{Caciagli}},
  \bibinfo{author}{\bibfnamefont{R.~J.} \bibnamefont{Baars}},
  \bibinfo{author}{\bibfnamefont{A.~P.} \bibnamefont{Philipse}},
  \bibnamefont{and} \bibinfo{author}{\bibfnamefont{B.~W.}
  \bibnamefont{Kuipers}}, \bibinfo{journal}{Journal of Magnetism and Magnetic
  Materials} \textbf{\bibinfo{volume}{456}}, \bibinfo{pages}{423 }
  (\bibinfo{year}{2018}), ISSN \bibinfo{issn}{0304-8853},
  \urlprefix\url{http://www.sciencedirect.com/science/article/pii/S0304885317334662}.

\bibitem[{\citenamefont{Wysin}(2012)}]{Wysin_web}
\bibinfo{author}{\bibfnamefont{G.}~\bibnamefont{Wysin}},
  \emph{\bibinfo{title}{Demagnetization fields}} (\bibinfo{year}{2012}),
  \urlprefix\url{https://www.phys.ksu.edu/personal/wysin/notes/demag.pdf}.

\bibitem[{\citenamefont{Franco et~al.}(2014)\citenamefont{Franco, D{\'e}jardin,
  and Kachkachi}}]{francoetal14jap}
\bibinfo{author}{\bibfnamefont{A.~F.} \bibnamefont{Franco}},
  \bibinfo{author}{\bibfnamefont{J.~L.} \bibnamefont{D{\'e}jardin}},
  \bibnamefont{and}
  \bibinfo{author}{\bibfnamefont{H.}~\bibnamefont{Kachkachi}},
  \bibinfo{journal}{Journal of Applied Physics} \textbf{\bibinfo{volume}{116}},
  \bibinfo{pages}{243905} (\bibinfo{year}{2014}),
  \eprint{https://doi.org/10.1063/1.4904750},
  \urlprefix\url{https://doi.org/10.1063/1.4904750}.

\bibitem[{\citenamefont{Var{\'o}n et~al.}(2013)\citenamefont{Var{\'o}n,
  Beleggia, Kasama, Harrison, Dunin-Borkowski, Puntes, and
  Frandsen}}]{Varon_dipolar_scirep2013}
\bibinfo{author}{\bibfnamefont{M.}~\bibnamefont{Var{\'o}n}},
  \bibinfo{author}{\bibfnamefont{M.}~\bibnamefont{Beleggia}},
  \bibinfo{author}{\bibfnamefont{T.}~\bibnamefont{Kasama}},
  \bibinfo{author}{\bibfnamefont{R.}~\bibnamefont{Harrison}},
  \bibinfo{author}{\bibfnamefont{R.}~\bibnamefont{Dunin-Borkowski}},
  \bibinfo{author}{\bibfnamefont{V.}~\bibnamefont{Puntes}}, \bibnamefont{and}
  \bibinfo{author}{\bibfnamefont{C.}~\bibnamefont{Frandsen}},
  \bibinfo{journal}{Scientific Reports} \textbf{\bibinfo{volume}{3}},
  \bibinfo{pages}{1234} (\bibinfo{year}{2013}),
  \urlprefix\url{http://dx.doi.org/10.1038/srep01234}.

\bibitem[{\citenamefont{Luttinger and Tisza}(1946)}]{luttis46pr}
\bibinfo{author}{\bibfnamefont{J.~M.} \bibnamefont{Luttinger}}
  \bibnamefont{and} \bibinfo{author}{\bibfnamefont{L.}~\bibnamefont{Tisza}},
  \bibinfo{journal}{Phys. Rev.} \textbf{\bibinfo{volume}{70}},
  \bibinfo{pages}{954} (\bibinfo{year}{1946}),
  \urlprefix\url{https://link.aps.org/doi/10.1103/PhysRev.70.954}.

\bibitem[{\citenamefont{Bastardis et~al.}(2017)\citenamefont{Bastardis, Vernay,
  Garanin, and Kachkachi}}]{BastardisEtal-jpcm2017}
\bibinfo{author}{\bibfnamefont{R.}~\bibnamefont{Bastardis}},
  \bibinfo{author}{\bibfnamefont{F.}~\bibnamefont{Vernay}},
  \bibinfo{author}{\bibfnamefont{D.-A.} \bibnamefont{Garanin}},
  \bibnamefont{and}
  \bibinfo{author}{\bibfnamefont{H.}~\bibnamefont{Kachkachi}},
  \bibinfo{journal}{Journal of Physics: Condensed Matter}
  \textbf{\bibinfo{volume}{29}}, \bibinfo{pages}{025801}
  (\bibinfo{year}{2017}),
  \urlprefix\url{http://stacks.iop.org/0953-8984/29/i=2/a=025801}.

\bibitem[{\citenamefont{Pigeau et~al.}(2012)\citenamefont{Pigeau, Hahn,
  de~Loubens, Naletov, Klein, Mitsuzuka, Lacour, Hehn, Andrieu, and
  Montaigne}}]{pigeauetal12prl}
\bibinfo{author}{\bibfnamefont{B.}~\bibnamefont{Pigeau}},
  \bibinfo{author}{\bibfnamefont{C.}~\bibnamefont{Hahn}},
  \bibinfo{author}{\bibfnamefont{G.}~\bibnamefont{de~Loubens}},
  \bibinfo{author}{\bibfnamefont{V.~V.} \bibnamefont{Naletov}},
  \bibinfo{author}{\bibfnamefont{O.}~\bibnamefont{Klein}},
  \bibinfo{author}{\bibfnamefont{K.}~\bibnamefont{Mitsuzuka}},
  \bibinfo{author}{\bibfnamefont{D.}~\bibnamefont{Lacour}},
  \bibinfo{author}{\bibfnamefont{M.}~\bibnamefont{Hehn}},
  \bibinfo{author}{\bibfnamefont{S.}~\bibnamefont{Andrieu}}, \bibnamefont{and}
  \bibinfo{author}{\bibfnamefont{F.}~\bibnamefont{Montaigne}},
  \bibinfo{journal}{Phys. Rev. Lett.} \textbf{\bibinfo{volume}{109}},
  \bibinfo{pages}{247602} (\bibinfo{year}{2012}),
  \urlprefix\url{https://link.aps.org/doi/10.1103/PhysRevLett.109.247602}.

\bibitem[{\citenamefont{J\"onsson and
  Garc\'{\i}a-Palacios}(2001)}]{jongar01prb}
\bibinfo{author}{\bibfnamefont{P.~E.} \bibnamefont{J\"onsson}}
  \bibnamefont{and} \bibinfo{author}{\bibfnamefont{J.~L.}
  \bibnamefont{Garc\'{\i}a-Palacios}}, \bibinfo{journal}{Phys. Rev. B}
  \textbf{\bibinfo{volume}{64}}, \bibinfo{pages}{174416}
  (\bibinfo{year}{2001}),
  \urlprefix\url{https://link.aps.org/doi/10.1103/PhysRevB.64.174416}.

\bibitem[{\citenamefont{Issa et~al.}(2013)\citenamefont{Issa, Obaidat, Albiss,
  and Haik}}]{issaetal13imsc}
\bibinfo{author}{\bibfnamefont{B.}~\bibnamefont{Issa}},
  \bibinfo{author}{\bibfnamefont{I.}~\bibnamefont{Obaidat}},
  \bibinfo{author}{\bibfnamefont{B.}~\bibnamefont{Albiss}}, \bibnamefont{and}
  \bibinfo{author}{\bibfnamefont{Y.}~\bibnamefont{Haik}},
  \bibinfo{journal}{International Journal of Molecular Sciences}
  \textbf{\bibinfo{volume}{14}}, \bibinfo{pages}{21266} (\bibinfo{year}{2013}),
  ISSN \bibinfo{issn}{1422-0067},
  \urlprefix\url{http://dx.doi.org/10.3390/ijms141121266}.

\bibitem[{\citenamefont{Rohart et~al.}(2007)\citenamefont{Rohart, Repain,
  Thiaville, and Rousset}}]{rohartetal07prb}
\bibinfo{author}{\bibfnamefont{S.}~\bibnamefont{Rohart}},
  \bibinfo{author}{\bibfnamefont{V.}~\bibnamefont{Repain}},
  \bibinfo{author}{\bibfnamefont{A.}~\bibnamefont{Thiaville}},
  \bibnamefont{and} \bibinfo{author}{\bibfnamefont{S.}~\bibnamefont{Rousset}},
  \bibinfo{journal}{Phys. Rev. B} \textbf{\bibinfo{volume}{76}},
  \bibinfo{pages}{104401} (\bibinfo{year}{2007}),
  \urlprefix\url{https://link.aps.org/doi/10.1103/PhysRevB.76.104401}.

\bibitem[{\citenamefont{Franco and Landeros}(2016)}]{fralan16jpdap}
\bibinfo{author}{\bibfnamefont{A.~F.} \bibnamefont{Franco}} \bibnamefont{and}
  \bibinfo{author}{\bibfnamefont{P.}~\bibnamefont{Landeros}},
  \bibinfo{journal}{Journal of Physics D: Applied Physics}
  \textbf{\bibinfo{volume}{49}}, \bibinfo{pages}{385003}
  (\bibinfo{year}{2016}),
  \urlprefix\url{http://stacks.iop.org/0022-3727/49/i=38/a=385003}.

\bibitem[{\citenamefont{Sabsabi et~al.}(2013)\citenamefont{Sabsabi, Vernay,
  Iglesias, and Kachkachi}}]{sabsabietal13prb}
\bibinfo{author}{\bibfnamefont{Z.}~\bibnamefont{Sabsabi}},
  \bibinfo{author}{\bibfnamefont{F.}~\bibnamefont{Vernay}},
  \bibinfo{author}{\bibfnamefont{O.}~\bibnamefont{Iglesias}}, \bibnamefont{and}
  \bibinfo{author}{\bibfnamefont{H.}~\bibnamefont{Kachkachi}},
  \bibinfo{journal}{Phys. Rev. B} \textbf{\bibinfo{volume}{88}},
  \bibinfo{pages}{104424} (\bibinfo{year}{2013}),
  \urlprefix\url{https://link.aps.org/doi/10.1103/PhysRevB.88.104424}.

\bibitem[{\citenamefont{Vernay et~al.}(2014)\citenamefont{Vernay, Sabsabi, and
  Kachkachi}}]{vernayetal14prb}
\bibinfo{author}{\bibfnamefont{F.}~\bibnamefont{Vernay}},
  \bibinfo{author}{\bibfnamefont{Z.}~\bibnamefont{Sabsabi}}, \bibnamefont{and}
  \bibinfo{author}{\bibfnamefont{H.}~\bibnamefont{Kachkachi}},
  \bibinfo{journal}{Phys. Rev. B} \textbf{\bibinfo{volume}{90}},
  \bibinfo{pages}{094416} (\bibinfo{year}{2014}),
  \urlprefix\url{https://link.aps.org/doi/10.1103/PhysRevB.90.094416}.

\end{thebibliography}

\end{document}